\documentclass[twocolumn,appendixfloats,tighten,numberedappendix]{openjournal}
\usepackage{amsmath}
\usepackage{graphicx}
\usepackage[T1]{fontenc} % if needed
\usepackage{fontawesome5}

\newcommand{\bq}{\boldsymbol q}
\newcommand{\bx}{\boldsymbol x}
\newcommand{\br}{\boldsymbol r}
\newcommand{\bp}{\boldsymbol p}
\newcommand{\bk}{\boldsymbol k}
\newcommand{\bPsi}{\boldsymbol{\Psi}}

\newcommand{\ihmpc}{\,h{\rm Mpc}^{-1}}

\newcommand{\quijote}{\texttt{Quijote} }

\newcommand{\kmax}{k_{\rm max}}
\newcommand{\ksph}{k_{\rm sph}}

\usepackage{xcolor}
\usepackage{textgreek}
\usepackage[utf8]{inputenc}
\usepackage[english]{babel}

\usepackage{hyperref}
\hypersetup{
    unicode, 
    colorlinks=true,
    linkcolor=linkcolor,
    citecolor=linkcolor,
    filecolor=linkcolor,
    urlcolor=linkcolor,
}
\usepackage{cleveref}
\usepackage{color,colortbl}
\definecolor{linkcolor}{rgb}{0.0,0.3,0.5}
\usepackage{tensind}
\tensordelimiter{?}
\DeclareGraphicsExtensions{.bmp,.png,.jpg,.pdf}
\usepackage{verbatim}
\usepackage[normalem]{ulem}
\usepackage{orcidlink}
\usepackage{soul}
\begin{document}

\title{\vspace{-0.8cm}\boldmath Control variates from Eulerian and Lagrangian perturbation theory:\\ Application to the bispectrum\vspace{-1.5cm}}
\author{Nickolas Kokron\textsuperscript{\faCookieBite}\orcidlink{0000-0002-5808-4708}}
\author{Shi-Fan Chen\textsuperscript{\faCookieBite,\faChessKing,\faWineGlass}\orcidlink{0000-0002-5762-6405}}
\email{kokron@ias.edu}
\email{sc5888@columbia.edu}

\affiliation{\textsuperscript{\faCookieBite}School of Natural Sciences, Institute for Advanced Study, 1 Einstein Drive, Princeton, NJ 08540, USA}
\affiliation{\textsuperscript{\faChessKing}Department of Physics, Columbia University, New York, NY, USA 10027}
\affiliation{\textsuperscript{\faWineGlass}NASA Hubble Fellowship Program, Einstein Fellow}

\begin{abstract}Control variates have seen recent interest as a powerful technique to reduce the variance of summary statistics measured from costly cosmological $N$-body simulations. Of particular interest are the class of control variates which are analytically calculable, such as the recently introduced `Zeldovich control variates' for the power spectrum of matter and biased tracers. In this work we present the construction of perturbative control variates in Eulerian and Lagrangian perturbation theory, and adopt the matter bispectrum as a case study. Eulerian control variates are analytically tractable for all $n$-point functions, but we show that their correlation with the $N$-body $n$-point function decays at a rate proportional to the sum-of-squared wavenumbers, hampering their utility. We show that the Zeldovich approximation, while possessing an analytically calculable bispectrum, is less correlated at low-$k$ than its Eulerian counterpart. We introduce an alternative -- the `shifted control variate' -- which can be constructed to have the correct tree-level $n$-point function, is Zeldovich-resummed, and in principle has an analytically tractable bispectrum. We find that applying this shifted control variate to the $z=0.5$ matter bispectrum is equivalent to averaging over $10^4$ simulations for the lowest-$k$ triangles considered. With a single $V=1({\rm Gpc}/h)^3$ $N$-body simulation, for a binning scheme with $N\approx 1400$ triangles from $k_{\rm min} = 0.04 \ihmpc$ to $k_{\rm \max} = 0.47  \ihmpc$, we obtain sub-2\% precision for every triangle configuration measured. This work enables the development of accurate bispectrum emulators -- a probe of cosmology well-suited to simulation-based modeling -- and lays the theoretical groundwork to extend control variates for the entire $n$-point hierarchy. 
\end{abstract}

\maketitle

\section{Introduction}
\label{sec:intro}
The advent of large-scale galaxy surveys has delivered precise maps of the Universe across many redshifts, revealing the non-Gaussian structure of the cosmic web at high significance. While most of the cosmological information in a galaxy survey is contained within its two-point statistics, higher-order statistics are complementary summaries which aid in constraining the fundamental parameters of the Universe by nature of being sensitive to different combinations than two-point summaries~\citep{2ptcollaboration2024parametermaskedmockdatachallenge}. While there are many non-Gaussian summary statistics considered, perhaps the most natural is the bispectrum -- the extension of the power spectrum to three powers of the density field -- which characterizes the skewness of the cosmic matter field in question~\citep{Scoccimarro_2000}. \par 
Despite its conceptual simplicity, modeling and analyzing the bispectrum is significantly more challenging than its two-point counterpart. On the modeling side, analytic tools such as perturbation theories of large-scale structure struggle to accurately reproduce the bispectrum with the exception of triangle configurations with very small wavenumbers (although see recent developments at one loop~\citep{Angulo_2015, Philcox_2022,D_Amico_2024, bakx2025oneloopgalaxybispectrumconsistent}). On the analysis side, measuring the bispectrum in a survey at high significance is computationally challenging. Significant care must be taken to characterize observational effects such as survey geometry~\citep{Pardede_2022, Wang_2025}, fiber collisions~\citep[in the case of spectroscopic surveys]{Hahn_2017, chudaykin2025reanalyzingdesidr11}, and even a proper characterization of the covariance of the bispectrum in an ideal simulation box is not trivial~\citep{Biagetti_2022}. \par 
These challenges in analytically modeling the bispectrum have led to significant interest in simulation-based models. $N$-body simulations of large-scale structure solve for the nonlinear and non-Gaussian distribution of dark matter down to substantially smaller scales than what is analytically accessible using perturbative techniques~\citep{Angulo_2022}. Measurements of the bispectrum in simulations -- significantly easier than when presented with observational challenges -- can be used as models for the signals measured in galaxy surveys. By performing $N$-body simulations at various points in cosmological parameter space, surrogate models of the bispectrum can be built which can smoothly interpolate its signals in regions where simulations were not performed (such as the BiHaloFit model of \cite{Takahashi_2020}).\par 
The high precision at which the bispectrum of matter and galaxies will be measured with stage-IV surveys requires that emulators of it must be highly accurate. This is difficult to achieve without averaging over many realizations at a fixed point in parameter space, hindering the efficacy of the emulation program -- simulations suffer from so-called `cosmic variance' due to the randomness inherent in their initial conditions. This issue with simulation-based modeling is not inherent to the bispectrum, and indeed extends itself to all summary statistics measured from simulations. \par 
To overcome these limitations, several techniques have been proposed to suppress the variance inherent in $N$-body simulations. The technique of `pairing and fixing'~\citep{Pontzen_2016,Angulo:2016hjd} is successful for the power spectrum but less-so for the simulation-based bispectrum and biased-tracer power spectra~\citep{Maion22}. The technique of control variates, a variance reduction tool widely adopted in Statistics~\citep{mcbook}, has seen a particular interest in the context of cosmology~\citep{chartier2020}. By exploiting a cheaper-yet-correlated simulation (with shared initial conditions), the method of control variates can be used to reduce the sample variance of simulation-based observables. These summary statistics measured from cheap simulations can be categorized into two classes: cheap simulations whose mean summary statistics, $\mu_c$, are known analytically~\citep{tassev2012,Kokron_2022, DeRose_2023,Hadzhiyska_2023} and those where the mean has to be determined from an ensemble average~\citep{chartier2021, Chartier:2022kjz, Ding:2022ydj, ding2025suppressingsamplevariancedesilike}. The latter case requires a careful study of the convergence and uncertainty on the mean, or else the total amount of variance reduction achievable is difficult to determine. \par
\begin{figure*}
    \centering
    \includegraphics[width=\linewidth]{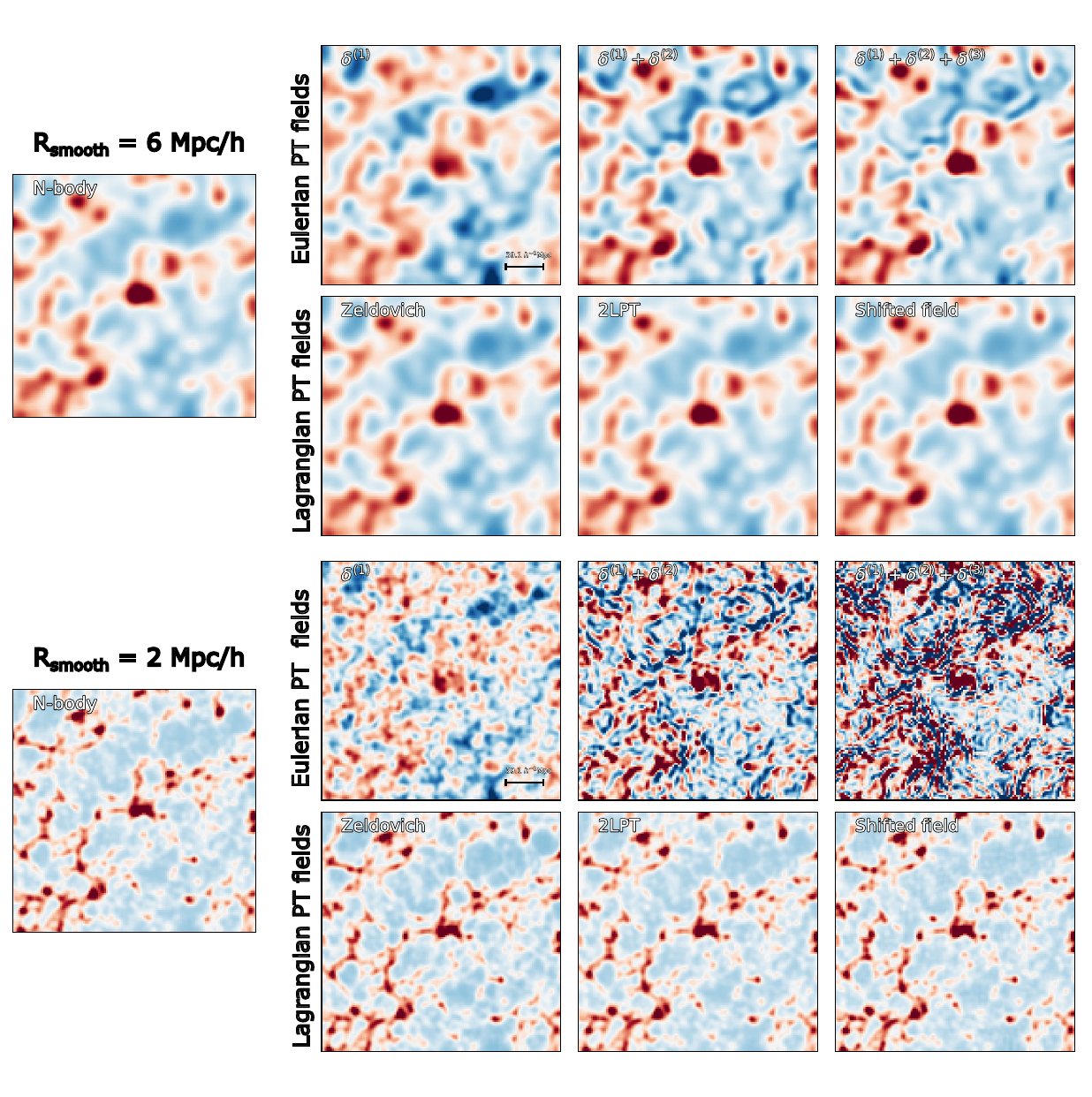}
    \caption{Visualization of the Eulerian and Lagrangian perturbative control variate fields considered in this work, as a function of smoothing scale. The left panels show the $N$-body density field centered around the largest overdensity of the simulation at $z=0.5$, projected across 20 Mpc/$h$. When filtered on a smoothing scale with $R_{\rm smooth} > \Sigma \sim 4.5 {\rm Mpc}/h$ (at which EPT decorrelates) the $N$-body, Eulerian and Lagrangian fields are all visually correlated. However, smoothing at a smaller scale reveals the rapid decorrelation of EPT, while the Lagrangian fields maintain a large degree of similarity to the $N$-body distribution. The color scale is chosen to be symmetric around $\pm 3 \sigma_{\rm lin}(R)$, where $\sigma_{\rm lin}(R)$ is the standard deviation of the Gaussian panel at that smoothing scale. }
    \label{fig:fieldplot}
\end{figure*}

 
The purpose of this paper is to carry out an in-depth investigation into the former class of models, which we dub \emph{perturbative} control variates, focusing on the application to the matter bispectrum. Both Eulerian and Lagrangian perturbation theories of structure formation can be used to create cheap-yet-correlated simulations of the late redshift Universe whose summary statistics are analytically understandable. The Eulerian approach will be shown to have the advantage of being simpler to compute analytically and a well-defined order-by-order expansion for the control variate exists. However, the Eulerian theory will pay the cost of being exponentially decorrelated after a scale $\Sigma$ (\cref{eqn:sigmadisp}) characterizing the average dispersion of the motions of galaxies in linear theory. We will then turn to the Lagrangian theory, where the first-order solution ~\citep[also known as the Zeldovich approximation]{1970A&A.....5...84Z} has an analytically tractable bispectrum. We also study the bispectrum in second-order Lagrangian Perturbation Theory (2LPT), which contains the correct tree-order bispectrum but loses analytic tractability. Finally, we introduce a class of hybrid control variate builts from the `shifted operator' basis of \cite{schmittfull2020modeling}, and show it possesses the correct tree-level $N$-point function while still being analytically resummable, outperforming both models. A visual summary of the different fields considered in this work is shown in Fig.~\ref{fig:fieldplot}, smoothed on two different scales with a Gaussian filter. The scales are chosen to be larger and smaller than the smoothing scale $\Sigma$ at which Eulerian fields decorrelate from the $N$-body density. \par
This paper is structured as follows: in \S~\ref{sec:toymodel} we explore creating a control variate for the bispectrum in a toy model where the non-linear field is generated by the Zeldovich approximation. In this toy model the `Eulerian' control variates are significantly simplified, and we can explain the structure of correlations between the Zeldovich density field and order-by-order Eulerian fields, as well as the correlation of the underlying bispectra. In \S~\ref{sec:nbodycv} we turn to control variates applied to the full $N$-body problem. We show that the Eulerian intuition developed in \S~\ref{sec:toymodel} holds in the $N$-body case and study the performance of Eulerian bispectra up to fifth order in the density field. We then turn to bispectrum control variates in LPT, studying the Zeldovich and 2LPT bispectra and how they correlate with the $N$-body result. We also show how to analytically match both Eulerian and the Zeldovich control variate to lattice-based realizations at high accuracy. This section concludes with an introduction of the `shifted control variate', which mixes desirable properties of both Eulerian and Lagrangian schemes. In \S~\ref{sec:results} we quantify the variance reduction achieved by each control variate as a function of scale, and compare these results to some of the past literature. We conclude with summarizing remarks and some future directions to be explored in \S~\ref{sec:conclusions}. 

\subsection{Conventions:}
All spectra measured in this work use the fiducial \quijote simulations~\citep{Villaescusa_Navarro_2020}, which have $N=(512)^3$ particles in boxes of size $L = 1\,h^{-1} {\rm Gpc}$. Bispectra are measured using \texttt{PolyBin3D}~\citep{philcox2024polybin3dsuiteoptimalefficient} whose real-space bispectrum estimator is an implementation of the estimator presented in \cite{Scoccimarro_1999, Sefusatti_2016}. The statistical properties of these bispectra come from an ensemble of the first $N=1000$ simulations in the fiducial \quijote suite, using the snapshot at $z=0.5$. There are two binning schemes adopted, in order to highlight different aspects of our control variates. The toy model of \S~\ref{sec:zerocv} where we treat the Zel'dovich approximation as the fully non-linear density field uses triangles defined by bins of width $\Delta k = 2k_f \approx 0.0125 \ihmpc$ until $\kmax = 0.15 \ihmpc$. This corresponds to $N=236$ bins. This is called the \textbf{Eulerian binning scheme} in the text.\par 
Analyses where the non-linear density field is the full $N$-body density field use a different binning scheme which has been tailored to simultaneously keep $\Delta \log k$ constant (and small), while extending to a higher $\kmax$ in order to capture more decorrelation, and ensuring a manageable number of triangles are included in the measurement. {We call this the \textbf{Zeldovich binning scheme}, which is composed of the union of two sets of bins. These bins are:}
\begin{itemize}
    \item Linearly-spaced $k$-bins with width $\Delta k = 3k_f$ until $k = 0.3 \ihmpc$. 
    \item Log-spaced $k$-bins between $0.3 \leq k/(\ihmpc) \leq 0.5$ with width $\Delta \ln k = 0.06$.
\end{itemize}
{The resulting triangles are computed from this full set of bins}. This results in $N=1434$ triangles. While $N_{\rm sims} < N_{\rm tri}$, we note that since we only concern ourselves with diagonal uncertainties or the diagonal part of the cross-correlation matrix, we are able to measure these at reasonably high statistical significance. We do not have sufficient statistics to resolve the full structure of the covariance matrix for this binning scheme. \par 

\section{Warm-up: bispectrum control variates in a toy model}
\label{sec:toymodel}
Consider the control variates problem applied to the real-space bispectrum, where we construct the random variable $y$
\begin{equation}
    y(k_1, k_2, k_3) = \hat{B}^{\rm sim}- \beta \left [ \hat{B}^{\rm CV} - \bar{B}^{\rm CV}\right ].
\end{equation}
$\beta$ is a Lagrange multiplier, $\hat{B}^{\rm sim}$ is the bispectrum measured in the $N$-body simulation, $\hat{B}^{\rm CV}$ is the control variate bispectrum measure from the cheap-yet-correlated simulation, $\bar{B}^{\rm CV}$ is the mean of the bispectrum, and we have suppressed the $(k_1, k_2, k_3)$ dependence on the right-hand side. Note that $\langle y \rangle = \langle \hat{B}^{\rm sim} \rangle$, and so $y$ is an un-biased estimator of the simulation bispectrum. Considering the univariate problem of minimizing the triangle-by-triangle uncertainty on the bispectrum, it is known that for the optimal $\beta$ that minimizes $\sigma^2_y$, the achieved variance suppression will be given by
\begin{equation}
\label{eqn:varredux}
   \frac{{\rm Var}[y]}{{\rm Var}[\hat{B}^{\rm sim}]} = 1 - \frac{{\rm Cov}^2[\hat{B}^{\rm sim}, \hat{B}^{\rm CV}]}{{\rm Var}[\hat{B}^{\rm sim}]{\rm Var}[\hat{B}^{\rm CV}]} = 1 - \rho^2_{\rm CV,sim},
\end{equation}
with $\rho_{\rm CV,sim}$ the bispectrum cross-correlation coefficient. A highly correlated control variate has the potential to substantially reduce the uncertainty in simulation-based bispectrum estimation, as has been achieved with the power spectrum. However, \cref{eqn:varredux} hinges on there being \emph{no uncertainty} associated with the estimation of the mean $\bar{B}^{\rm CV}$. In the presence of uncertainty in estimate of the mean, when $N$ independent surrogate simulations are used to estimate $\bar{B}^{\rm CV}$ the uncertainty will increase by 
\begin{equation}
    \sigma_{y}^2 \to \sigma_y^2 + \beta^2 \frac{\sigma_c^2}{N},  
\end{equation}
Unless it holds that $\beta^2 \sigma_c^2 / N \ll \sigma_y^2 \approx \sigma_x^2 (1 - \rho_{x,c}^2)$, any potential gains from the control variate's correlation will be dwarfed by uncertainty on its mean estimate. \par 
We can analytically estimate how large $N$ has to be in order to not dilute the gain in precision. We define $r$ to the ratio of variances with and without an empirically determined mean for the control variate
\begin{align}
    r &= 1+\frac{\beta^2 \sigma_c^2}{N\sigma_x^2 (1 - \rho_{x,c}^2) },
\end{align}
where we have switched to a shorthand notation where $x$ is $\hat{B}^{\rm sim}$ and $c$ is the control variate $\hat{B}^{\rm CV}$. In the case of the optimal Lagrange multiplier, $\beta^*$, 
\begin{equation}
    \beta^* \equiv \frac{\sigma_{xc}}{\sigma_c^2}  = \rho_{xc}\frac{\sigma_x}{\sigma_c}, 
\end{equation}
and so we can recast $r$ in a way that depends solely on $\rho$
\begin{align}
\label{eqn:clearr}
    r &= 1 + \frac{\rho_{x,c}^2}{N ( 1 - \rho_{x,c}^2)}.
\end{align}
\cref{eqn:clearr} makes it clear what are the requirements imposed on a control variate if its mean is to be sampled empirically -- the more correlated the control variate, the more realizations of it will be needed to not have the uncertainty on the mean saturate the variance reduction. The variance reduction is halved when 
\begin{equation} 
\label{eqn:cvN}
N = 1/ (1-\rho^2) - 1. 
\end{equation}
In the regime of large cross-correlation, we see that $N$ is simply the ratio of the $N$-body variance to the control variate's variance -- also called the `volume multiplier'. Empirical control variates require the uncertainty on the mean to be reduced by a factor equivalent to the volume multiplier achieved by the control variate. For a volume multiplier of $10^4$, the equivalent number of surrogate simulations and bispectra measurements would have to be carried out. Emulation suites which sample the parameter space with $\mathcal{O}(100)$ simulations would require, then, $10^6$ simulation-bispectrum runs. \par 
It is evidently desirable that a control variate have its mean be determined as precisely as possible, and the most precision one can achieve is through an analytic calculation of its signal. {For an analytic control variate, only one surrogate simulation is needed for each expensive simulation, a drastic reduction compared to empirical control variates.} The suitability of different analytically calculable bispectra as a control variate is one of the main aims of this work. \par 
{In the remainder of this section, we will introduce an analytic toy model of bispectrum control variates where many quantities are calculable. This will help understand what to expect in the full $N$-body problem. Specifically, in \S~\ref{sec:zerocv} we start with the problem of a `Gaussian' control variate where the mean is zero, and use the Zeldovich approximation to generate the `true' non-linear density field. In \S~\ref{sec:ZAbk} we compute the Eulerian perturbation theory corrections to the analytic bispectrum control variate in this toy model.} 
\subsection{{Gaussian control variates}}
\label{sec:zerocv}
\begin{figure*}
\centering
    \includegraphics[width=\linewidth]{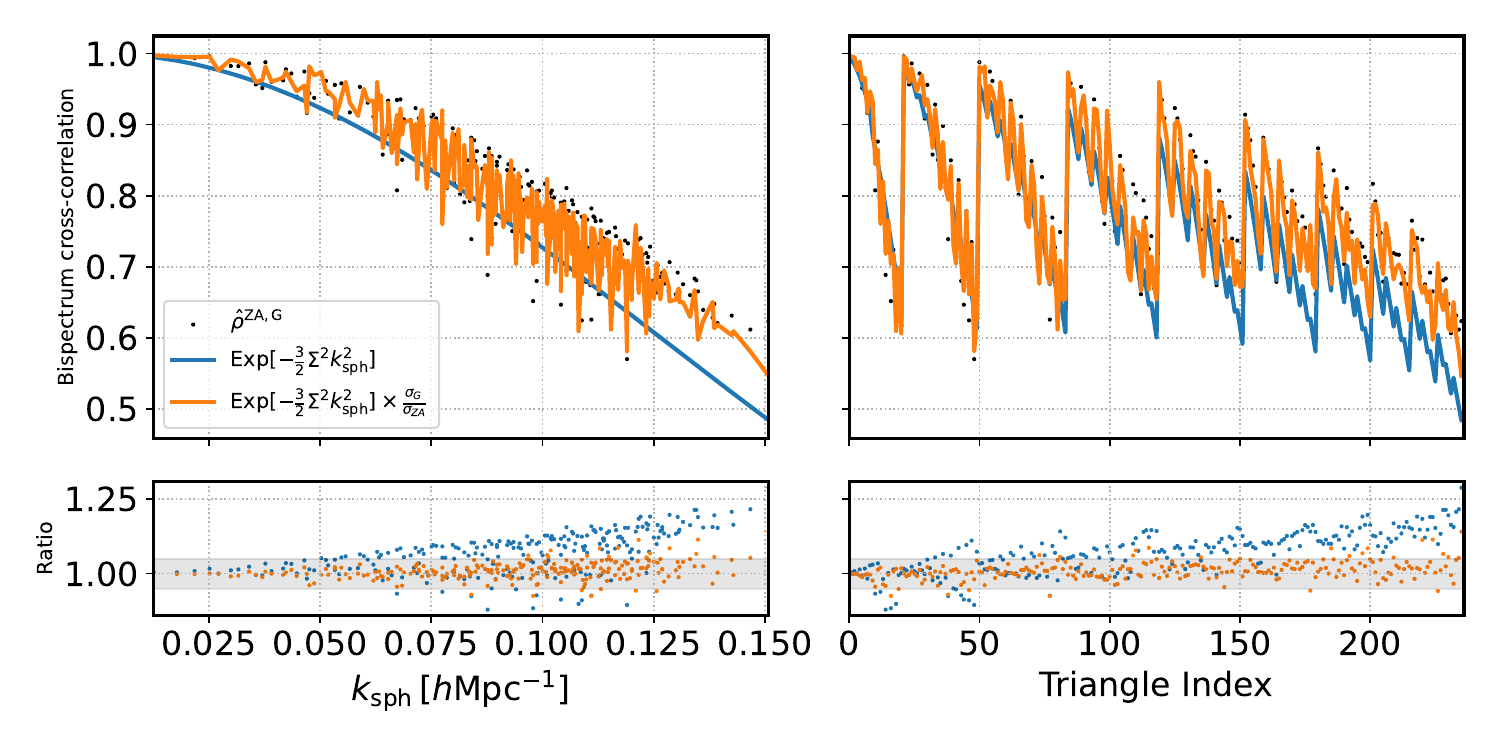}
    \caption{\emph{Top:} Cross-correlation coefficient between bispectra in Zeldovich boxes and those measured from the same Gaussian initial conditions. The shape of this cross-correlation coefficient is in close agreement with the prediction from \cref{eqn:rhozag}. \emph{Bottom:} Ratio of the empirically measured cross-correlation coefficients and the predictions from \cref{eqn:rhozagapprox}, in blue and \cref{eqn:rhozag}, in orange. The grey bands denote $\pm 5\%$ residuals. The left and right panels show the same data, but as a function of $\ksph = \sqrt{(k_1^2 + k_2^2 + k_3^2)/3}$ on left and the \emph{triangle index} representation on the right. The dominant behavior comes from the derived $\ksph$ dependence. The gray band in the residuals denotes $\pm 5\%$ deviations from unity.}
    \label{fig:rhozag}
\end{figure*}
While for the power spectrum dramatic reduction in variance can be achieved with a highly correlated observable with analytically tractable means (such as the power spectrum in the Zeldovich approximation)~\citep{Kokron_2022,DeRose_2023}, this is less feasible for higher $N$-point functions such as the bispectrum. Analytic predictions of higher $N$-point functions are more computationally intensive than the power spectrum: for example, the bispectrum in the Zeldovich approximation has recently been calculated by~\citep{chen2024bispectrumlagrangianperturbationtheory}, but evaluation times for a single triangle remain on the order of 1 second. Furthermore, in order to obtain sufficiently precise predictions of these N-point functions, it is important to take into account the discreteness of Fourier modes in simulations boxes beyond the usual continuous approximations for binning. Bispectrum estimators involve averaging over many triangle configurations that fall into a bin, with an analytic estimate for the number of triangles in a bin of width $\Delta k$, given by~\citep{Sefusatti10}
\begin{equation}
\label{eqn:ntri}
    N_{123} \approx 8\pi k_1 k_2 k_3 (\Delta k)^3 \frac{V^2}{(2\pi)^6}.
\end{equation}
The number of triangles is a rapidly growing function of $k$, and for volumes comparable to the fiducial simulation suites we use in this simulation, we must average over millions of configurations -- this naive average is unfeasible even with rapid evaluations on the order of a second. That said, much work has been done in the direction of simplifying this bin-averaging and we will return on approximations to improve the applicability of the ZA bispectrum in \S~\ref{subsubsec:analyticbk}. \par 
However, consider the bispectrum calculated from a linear, Gaussian density field which has seeded the full $N$-body simulation whose variance we wish to cancel. In this case, we can write an estimator for its \emph{linear} bispectrum
\begin{equation}
    \hat{B}^{111}(k_1, k_2, k_3 ) = \frac{1}{N_{123} V} \sum_{k_1, k_2, k_3} \delta_1 \delta_2 \delta_3 \delta^D_{123},
\end{equation}
where $N_{123}$ is the number of triangles in a given bin. For any given set of initial conditions, $\hat{B}^{111}$ doesn't have to be zero. At the same time, it's immediately clear that $\langle B^{111} \rangle = 0$. Nevertheless, this trivial bispectrum could still a useful control variate. This is because despite possessing a zero mean, its covariance is non-zero  
\begin{equation}
    {\rm Cov} [ \hat{B}^{\rm sim}, \hat{B}^{111} ] \neq 0.
\end{equation}
That this covariance is non-zero can be readily seen from considering the Gaussian disconnected contribution to the bispectrum covariance. Indeed, many analyses of the bispectrum rely on this Gaussian disconnected covariance in lieu of difficulties in estimating its full form. Thus, we expect the linear field's correlation with the $N$-body field to also be useful in computing a trivial control variate. \par 
Let's elucidate the structure of this zero-mean Gaussian control variate. Consider the following toy model: take the $N$-body field, $\delta^N$, to be the density field after being displaced by the Zeldovich approximation:
\begin{equation}
\label{eqn:zadens}
    \delta^N (\bk) = \int d^3q\, e^{i\bk\cdot \bq} \left [e^{i \bk \cdot \bPsi^{\rm ZA}(\bq)} - 1\right],
\end{equation}
where $\bPsi^{\rm ZA} (\bq)$ is the Zeldovich displacement (later defined in \cref{eqn:zadisp}). The Gaussian disconnected part of this covariance is given schematically\footnote{Neglecting geometric factors arising from the number of triangles or degeneracies depending on specific triangle shape.} by 
\begin{align}
    {\rm Cov} [ \hat{B}^{\rm sim}, \hat{B}^{111} ] &\propto \langle \delta_{1}^N\delta_{2}^N\delta_{3}^N | \delta_{4}\delta_5 \delta_6 \rangle, \\
\nonumber    &\approx \langle \delta^N \delta \rangle\langle \delta^N \delta \rangle\langle \delta^N \delta \rangle + \cdots  \\
\nonumber    &\propto e^{-\frac{1}{2} \Sigma^2 (k_1^2 + k_2^2 + k_3^2)} P_{\rm lin}(k_1) P_{\rm lin}(k_2) P_{\rm lin}(k_3),
\end{align}
where we have used $\langle \delta^N \delta\rangle'(k) = e^{-\frac{1}{2} \Sigma^2 k^2}P_{\rm lin} (k)$ and the displacement dispersion is
\begin{equation}
\label{eqn:sigmadisp}
    \Sigma^2 = \frac{1}{3} \langle |\Psi(\bq)|^2\rangle =  \frac{1}{6\pi^2} \int dk\, P_{\rm lin}(k) .
\end{equation}
This relation is exact in the Zeldovich approximation; it is also a good approximation in the general case given that most of the decorrelation between initial conditions and the final density field are due to the bulk linear motions \citep{Chisari19}.
From the above we can write the bispectrum cross-correlation coefficient as
\begin{equation}
\label{eqn:rhozag}
    \rho (k_1, k_2, k_3) = e^{-\frac{1}{2} \Sigma^2 (k_1^2 + k_2^2 + k_3^2)} \frac{\sigma_G}{\sigma_{ \rm ZA}},
\end{equation}
where $\sigma_{G, {\rm ZA}}$ is the standard deviation of the Gaussian / ZA bispectra respectively, for that given triangle configuration. If the Gaussian and ZA bispectra have covariances which are dominated by disconnected terms, and assuming that $P^{\rm ZA} \approx P_{\rm lin}$ at large scales (up to smoothing of any BAO wiggles) we find the cross-correlation coefficient between the two bispectra is given purely by the exponential damping term:
\begin{equation}
\label{eqn:rhozagapprox}
    \rho(k_1, k_2, k_3) \approx e^{-\frac{1}{2} \Sigma^2 (k_1^2 + k_2^2 + k_3^2)},
\end{equation}
and thus, using the Gaussian bispectrum of an $N$-body simulation as a control variate should provide variance cancellation on large scales compared to $\Sigma$. Notice that the decorrelation is exponential in the variable $k_1^2 + k_2^2 + k_3^2$. Thus, we will interchangeably show our results in terms of either \emph{triangle index}, $i$, or the `spherical wavenumber' 
\begin{equation*}
\ksph \equiv \sqrt{(k_1^2 + k_2^2 + k_3^2)/3}
\end{equation*}
defined in \cite{Tomlinson:2022xud}. Any scatter at fixed $\ksph$ indicates dependence on the `azimuthal' and `polar' angles of the spherical bispectrum ($\phi_{\rm sph} = \arctan(k_2 / k_1)$ and $\theta_{\rm sph} = \arctan ( \sqrt{k_1^2 + k_2^2}/k_3)$), which we expect to be sub-dominant when exploring the decorrelation for a control variate. 

As an explicit check of this decorrelation, we calculate the bispectrum in the Gaussian initial conditions of $N$-body simulations, as well as from the Zeldovich-displaced field seeded by these same initial conditions, at $z=0.5$. Computing these bispectra for $N=1000$ boxes we can measure the cross-correlation coefficient empirically, and this is shown in Fig.~\ref{fig:rhozag}. The curves in the figure also show the two approximations for this correlation coefficient, \cref{eqn:rhozagapprox} and \cref{eqn:rhozag}. The measured correlation is in close agreement with that of \cref{eqn:rhozag}, demonstrating that even a bispectrum with zero mean can serve as a control variate.\par 

The advantage of this Gaussian control variate is that one does not have to be concerned with subtle aspects of comparing a numerically-measured bispectrum to an analytic prediction, or ensuring the uncertainty on its mean is negligible. The mean, being zero, will remain zero even when considering averaging over various triangles within a bin. While clearly convenient, the utility of this Gaussian control variate is somewhat limited. The exponential decorrelation that affects `linear control variates' (discussed in \cite{Hadzhiyska_2023}) is even more dramatic here, since each leg of the triangle contributes its own suppression. Is it possible to do better while still retaining analytic control over the bispectrum?

\subsection{Perturbative control variates in the Zeldovich approximation}
\label{sec:ZAbk}
\begin{figure}
    \centering
    \includegraphics[width=.99\linewidth]{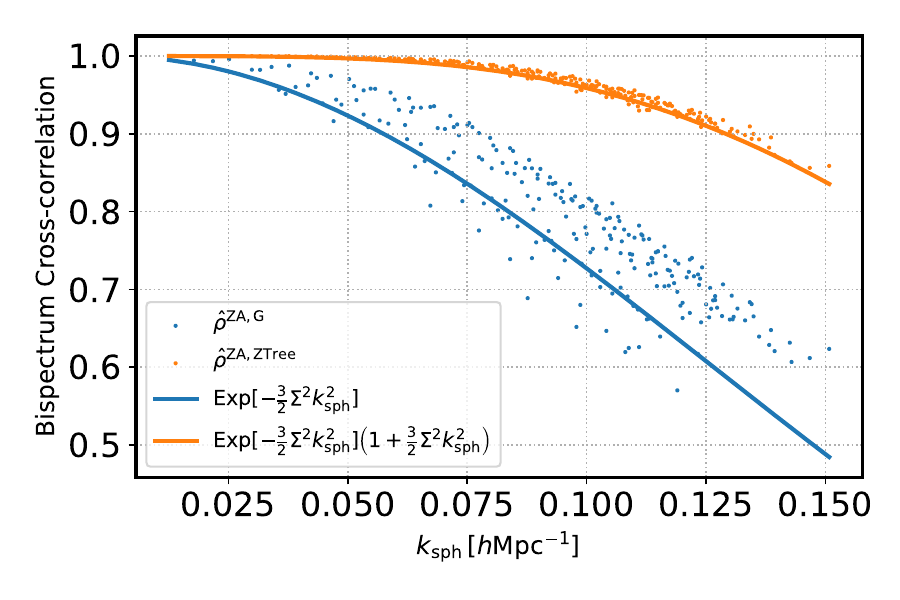}
    \caption{Bispectrum cross-correlation coefficient between the Zeldovich density field and Eulerian realizations, including the `Zeldovich tree level' bispectrum of \cref{eqn:bktreeza}. The cross-correlation coefficient is enhanced by a term $\propto \ksph^2,$ as argued analytically. We also see a significant tightening of the cross-correlation coefficients for different triangle configurations }
    \label{fig:za2bk}
\end{figure}
In this toy model where the `expensive' simulation is the Zeldovich density field, we can perturbatively expand \cref{eqn:zadens} to find `Eulerian' PT kernels for the Zeldovich approximation \citep{Grinstein87}:
\begin{align}
    \delta^N &\approx \int d^3q e^{i \bk \cdot \bq} \sum_n \frac{(i \bk \cdot \bPsi(\bq))^n}{n!} \\
    &\approx \sum_{n=1}^\infty \int_{\bk_1 \cdots \bk_n} \delta^{D} \left ( \bk - \sum_{i=1}^n \bk_i \right ) \underbrace{\frac{1}{n!}  \prod_{i=1}^n \frac{\bk \cdot \bk_i}{k_i^2}}_{\equiv  Z_n(\bk_1, \cdots, \bk_n)} \delta_{\bk_i},
\end{align}
The corresponding tree-level bispectrum is
\begin{equation}
\label{eqn:bktreeza}
    B^{211}_{\rm ZA}(\bk_1, \bk_2, \bk_3) = \frac{\bk_1 \cdot \bk_2}{k_2^2}\frac{\bk_1 \cdot \bk_3}{k_3^2} P(k_2) P(k_3) + {\rm cyclic}, 
\end{equation}
where $\bk_1 = -(\bk_2 + \bk_3)$. The disconected covariance of the tree-level bispectrum with the Zeldovich spectrum will involve contractions
\begin{align}
    {\rm Cov} [ \hat{B}^{\rm sim}, \hat{B}^{211} ] &\sim \langle \delta_{1}^N\delta_{2}^N\delta_{3}^N | \delta_1^{(2)} \delta_2 \delta_3 \rangle  \\
    &    \sim \langle \delta^N \delta^{(2)} \rangle \langle \delta^N \delta \rangle\langle \delta^N \delta \rangle + \cdots.
\end{align}
The inclusion of the tree-level bispectrum (beyond the Gaussian term) now has exponential suppression in two of the legs as well as a new correlator -- of the Zeldovich field with its second-order Eulerian counterpart, $\delta^{(2)}$. We can calculate this correlator exactly and see its effect on the cross-correlation. To see this, note that the second-order `Eulerian Zeldovich' density field, which is given by
\begin{align}
\label{eqn:za2}
    \delta^{(2)} (\bx) &= \frac{1}{2} \partial_{x,i} \partial_{x,j} [\Psi_i (\bx) \Psi_j (\bx)],
\end{align}
can be re-written as 
\begin{equation}
\label{eqn:za2field}
      \delta^{(2)} (\bx)  = \frac{2}{3} \delta_0^2 (\bx) - \Psi_i (\bx) \partial_i \delta_0 (\bx) + \frac{1}{2} s_0^2 (\bx),
\end{equation}
where $s_0^2$ is $s_{0,ij}s_{0,ij}$ and $s_{0,ij} = (\partial^{-2}\partial_i \partial_j - \delta_{ij}/3)\delta_0$ is linear tidal field tensor. Note that the initial fields and their derivatives are evaluated in Eulerian coordinates above. This expression possesses the same degrees of freedom as the standard $F_2 (\bk)$ kernel in Eulerian perturbation theory, but with a different weighting of the $\delta_0^2$ and $s_0^2$ coefficients. We write the $\langle \delta^{N} \delta^{(2)} \rangle$ power spectrum as a correlator in LPT
\begin{widetext}
\begin{equation}
    \langle \delta^{N} \delta^{(2)}\rangle' = \int d^3q e^{i \bk \cdot \textbf{r}}  \left \langle e^{i \bk \cdot \bPsi(\bq)} \left [ \frac{2}{3} \delta^2 (\bx) - \Psi_a (\bx) \partial_a \delta (\bx) + \frac{1}{2} s^2 (\bx) \right ] \right \rangle_{\textbf{r} = \bx - \bq}.
\end{equation}

Each of these terms can be evaluated exactly using the cumulant expansion theorem. The spectrum is given by\footnote{There are also contributions at zero-lag for individual correlators but they cancel when all terms are included.} 
\begin{align}
\label{eqn:nz2}
    \langle \delta^{N} \delta^{(2)}\rangle' = -e^{-\frac{1}{2} k^2 \Sigma^2} \int d^3q e^{i \bk \cdot \textbf{r}} k_i k_j \left [ \frac{2}{3} \langle \Psi_i \delta \rangle \langle \Psi_i \delta \rangle - \langle \partial_a \delta \Psi_i \rangle \langle \Psi_a \Psi_j \rangle + \frac{1}{2} \langle \Psi_i s_{ab} \rangle \langle \Psi_j s_{ab} \rangle \right ]
\end{align}
\end{widetext}
The Fourier transform in \cref{eqn:nz2} can be done to re-express the sub-spectra in the form of mode-coupling kernels in Eulerian PT; indeed, this yields the simple expression $\langle \delta^{N} \delta^{(2)} \rangle = \langle \delta^{(2)}\delta^{(2)} \rangle \exp(-k^2\Sigma^2/2)$. Taking the asymptotic limit of $P_{22}$ yields the IR contribution
\begin{equation}
    \label{eqn:kscaling}
    \langle \delta^{N} \delta^{(2)}\rangle'(k) \supset k^2\Sigma^2e^{-\frac{1}{2}k^2\Sigma^2} P_{\rm lin}(k).
\end{equation}
This IR contribution is partially canceled by the IR contribution to the damping exponent in $\langle N L^{(1)} \rangle$, leading to an overall IR contribution to the second-order field equal to 
\begin{equation}
    \left(1 + k^2 \Sigma^2 \right) e^{-\frac{1}{2}k^2\Sigma^2}  P_{\rm lin}(k) \approx \left( 1 + \frac12 k^2 \Sigma^2 \right) P_{\rm lin}(k)
\end{equation}
at large scales. This large-scale enhancement of the cross correlation is in turn canceled by the IR contributions to $P_{22}$ in the denominator when computing the correlation between the two fields, leading to damping due to bulk displacements beginning only at quartic order in $k$. In other words, while the cross-covariance between the fully nonlinear and quadratic fields are artificially enhanced by long-wavelength modes, their cross-correlation is suppressed by an asymptotically equally large enhancement in the variance of the quadratic field by those modes. Computing the matter field up to cubic order doesn't further reduce the leading-order $k^2$ decorrelation---since it is already zeroed in the quadratic cross correlation---but increases the correlation towards smaller scales due {to} better matching the mode coupling. Regardless, while the leading IR contributions in the form of polynomials of $k \Sigma$ cancel the decorrelation due to bulk modes to some extent, they are eventually overcome by the exponential in e.g. Equation~\ref{eqn:kscaling}. We thus see that including the tree-level control variate improves the correlation, but does not cancel the exponential decay at $k \gtrsim \Sigma^{-1}$. The interested reader is referred to Appendix~\ref{app:ptcorr} for more explicit calculations of the correlation coefficients and comparisons to simulations.

In Fig.~\ref{fig:za2bk} we show the cross-correlation coefficient between the full Zeldovich bispectrum and the `gaussian + tree-level' subset. We also plot the expected increase in correlation derived in \cref{eqn:kscaling}. The inclusion of the tree-level bispectrum boosts the cross-correlation coefficient from $\sim 60\%$ at $\ksph = \kmax = 0.15 \ihmpc$, to $\sim 85\%$, and the scale-dependence of the cross-correlation reflects the leading-order cancellation of the IR contribution derived above. We also observe the correlation coefficient is a very tight function of $\ksph$ when this additional term is included -- variations in covariance from non-equilateral triangles are significantly more well-captured with the inclusion of the tree level spectrum. \par 
Given the success of the tree-level Zeldovich bispectrum in this simplified toy model, we are motivated to investigate the performance of perturbative control variates in `real-world' applications where the $N$-body bispectrum is the object whose variance we wish to cancel. 
\section{Perturbative control variates and N-body simulations}
\label{sec:nbodycv}
In the previous section we showed, in an analytic toy model, that while Gaussian control variates (also referred to as `linear control variates' in the context of the power spectrum) can be beneficial, their variance suppression in a polyspectrum decays as an exponential in the sum-of-squared wavenumbers. However, we also saw that Eulerian perturbation theory provides an order-by-order method to construct a control variate which improves variance suppression to smaller scales. In this section we shall go beyond the Zeldovich approximation and use bispectra measured from $N$-body simulations as the object of interest. \par 
We will begin with a discussion of bispectra measured from Eulerian PT on the lattice, up to bispectra of order $\mathcal{O}(\delta^5)$. In \S~\ref{subsec:lagcv} we turn to bispectra in Lagrangian perturbation theory, where we evaluate the Zeldovich and 2LPT bispectra. This includes a discussion on how to compare grid-based and analytic Zeldovich bispectra to high accuracy. \S~\ref{subsec:shiftcv} introduces \emph{shifted control variates}, which use the shifted operator basis of \cite{schmittfull2020modeling} to construct a control variate which combines optimal characteristics of Eulerian and Lagrangian theory. 

\subsection{Eulerian Control Variates}
In Eulerian perturbation theory, a recursion relation can be used to generate the $n$-th order Eulerian density and velocity fields. This recursion relation is given in matrix form by~\citep{Bernardeau:2001qr, Taruya_2018}
\begin{widetext}
    \begin{equation}
\label{eqn:gridspt}
    \begin{pmatrix}
        \delta_n(\bx) \\\theta_n(\bx)
    \end{pmatrix} = \frac{2}{(2n+3)(n-1)} 
    \begin{pmatrix}
        n+1/2 & 1 \\
        3/2 & n
    \end{pmatrix} \sum_{m=1}^{n-1}
    \begin{pmatrix}
        (\nabla \delta_m) \cdot \boldsymbol{u}_{n-m} + \delta_m \theta_{n-m} \\
        [\partial_j (\boldsymbol{u}_m)_k][\partial_k (\boldsymbol{u}_{n-m})_k] + \boldsymbol{u}_m \cdot (\nabla \theta_{n-m})
    \end{pmatrix}.
\end{equation}
\end{widetext}
Given a linear density field $\delta_{\rm lin}(\bx) = \theta_{\rm lin}(\bx)$ that seeds an $N$-body simulation, evaluating this recursion relation produces field-level realizations of Eulerian PT.
\begin{figure}
    \centering
    \includegraphics[width=1.0\linewidth]{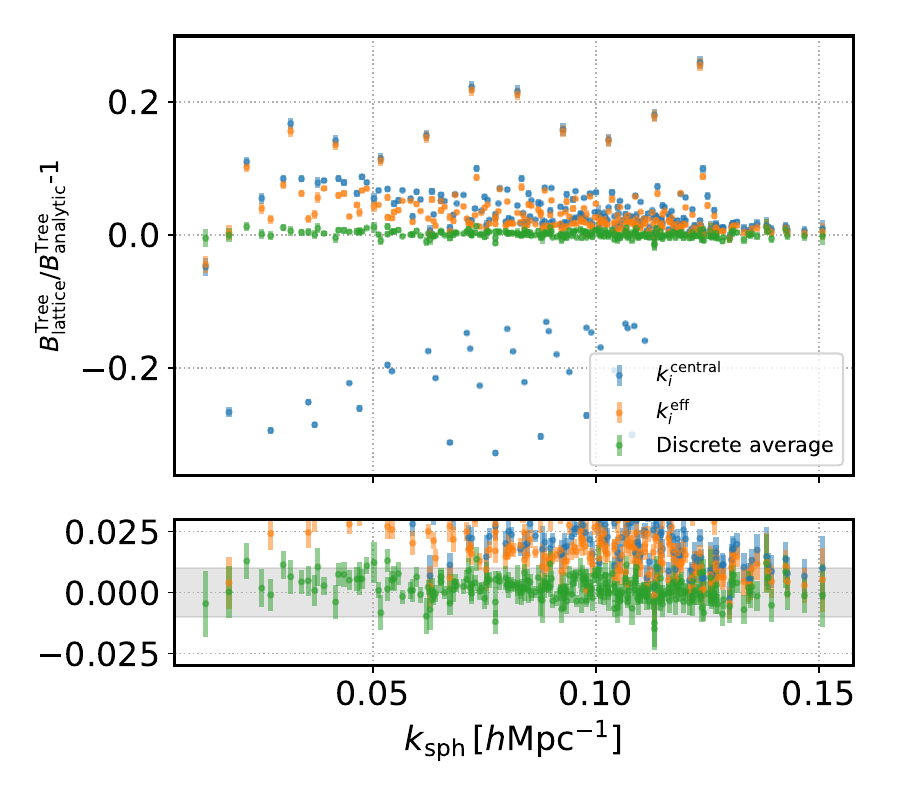}
    \caption{\emph{Top panel:} Relative deviation of tree level bispectrum averaged over $N=1000$ lattice-based realizations of the $\delta^{(2)}$ field compared to analytic predictions. The blue points show the tree level bispectrum when evaluated at the central $(k_1, k_2, k_3)$ of the bin, the orange curve shows the prediction when using the `effective triangle' in \cref{eqn:keff} and the green points show the result from averaging over all discrete triangles. The error induced due to not bin-averaging can reach 30\% for triangles in this scheme where $\Delta k = 2k_f$. \emph{Bottom panel:} The same panel, but zoomed to show the size of spread when the discrete bin average is made. The gray bands indicate 1\% scatter around zero. Error bars correspond to error bars on the mean of $N=1000$ realizations.}    \label{fig:btreelattice}
\end{figure}
\subsubsection{Tree-level Eulerian control variate}
\label{subsec:treecv}
\begin{figure*}
    \centering
    \includegraphics[width=\linewidth]{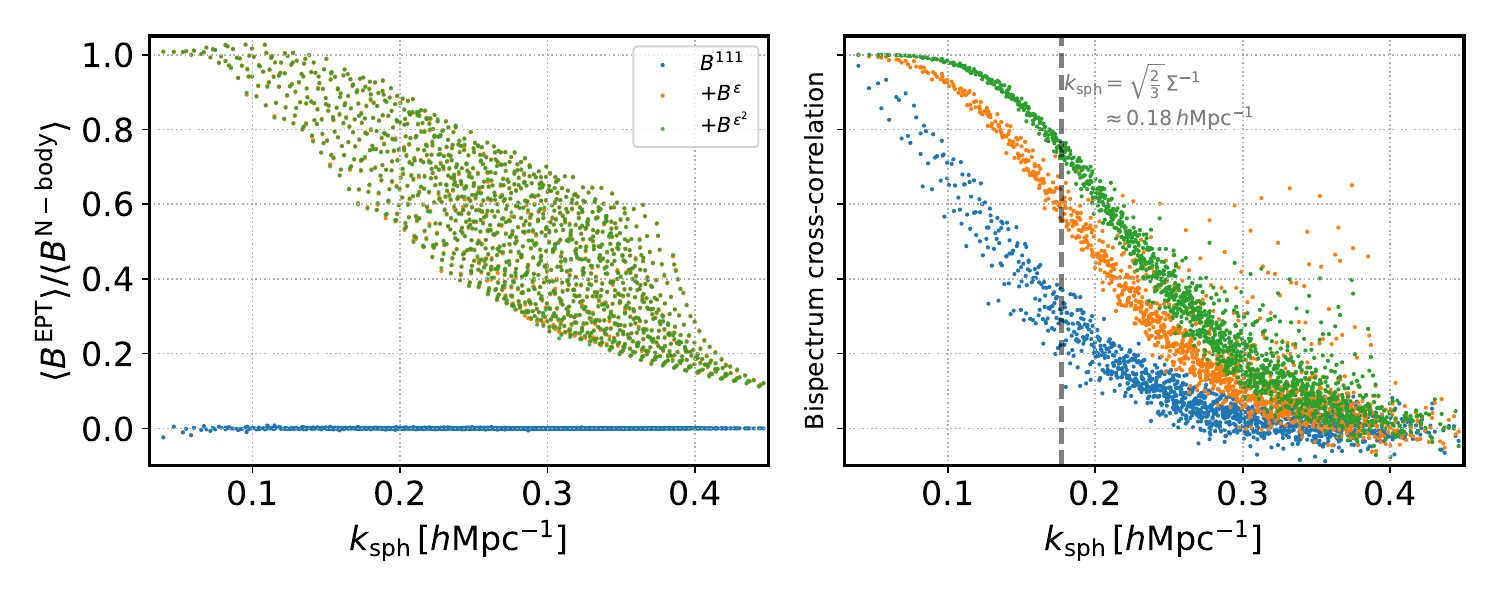}
    \caption{\emph{Left:} Ratio of bispectra in Eulerian perturbation theory, to order $\mathcal{O}(\epsilon^2)$ divided by the $N$-body bispectrum, as a function of $\ksph^2 = (k_1^2 + k_2^2 + k_3^2)/3$. Each bispectrum is averaged over $N=1000$ realizations. The blue points indicate the Gaussian bispectrum {(consistent with zero)}, orange points include the tree-level contribution, and the green points include terms up to the bispectrum diagrams $B^{221}$ and $B^{311}$, which have zero mean {and thus should not alter the plotted ratio}. \emph{Right:} Cross-correlation coefficients between the full $N$-body bispectrum and different realizations of the bispectrum measured in lattice Eulerian perturbation theory. Despite their zero-mean, the Gaussian term, as well as the fifth-order-in-$\delta$ terms, both contribute to increase the cross-correlation coefficient with the $N$-body bispectrum, and thus the total variance reduction achievable with EPT. The dashed vertical line in the right panel corresponds to the $\ksph$ mode at which one e-folding of decorrelation is expected, according to \cref{eqn:rhozagapprox}.}
    \label{fig:ecv}
\end{figure*}
The second-order Eulerian density field can be expressed as 
\begin{equation}
\label{eqn:ept2field}
    \delta^{(2)} (\bx) = \frac{17}{21} \delta^2 (\bx) - \bPsi(\bx) \cdot \nabla \delta (\bx) + \frac{2}{7} s^2 (\bx).
\end{equation}
This density field has the same analytic structure as the second-order ZA density field in \cref{eqn:za2}, with different coefficients for the growth and tidal contributions. We then measure the bispectrum of the full second-order density field 
\begin{equation}
\label{eqn:d1+d2}
    \delta(\bx) = \delta^{\rm lin} (\bx) + \delta^{(2)} (\bx).
\end{equation}

The bispectrum of the density field in \cref{eqn:d1+d2} is not an ideal control variate -- it contains a one-loop contribution from $B^{222} = \langle \delta^{(2)}_1\delta^{(2)}_2\delta^{(2)}_3\rangle$ -- which is difficult to evaluate analytically. {A detailed discussion for how we circumvent this loop diagram (and higher-order) contributions is detailed in Appendix~\ref{app:treebk}. In essence, we multiply the second-order density field by a counting parameter $\epsilon$, $\delta^{(2)} \to \epsilon \delta^{(2)}$. We then have that $B^{211} \sim \mathcal{O}(\epsilon)$, while $B^{222} \sim \mathcal{O}(\epsilon^3)$. A linear combination of bispectra from the Gaussian field and from this $\epsilon$-rescaled field is be used to extract the tree-level, $\mathcal{O}(\epsilon)$, piece at high numerical precision.}   \par 
The tree-level bispectrum, as measured in simulations, must also be computed to high accuracy in order to ensure the mean is unbiased. The leading theoretical uncertainty that has to be managed in order to achieve agreement between the tree level bispectrum measured on the lattice, and its analytic prediction, is to correct for the effect of the bin size that contributes to triangle estimation. In principle, for a given bin, the bispectrum measured in that bin is given by 
\begin{equation}
    B(k_1, k_2, k_3) = \frac{1}{N_{123}} \sum_{\bk_1, \bk_2, \bk_3 \in {\rm bin}} B^{\rm tree} (k'_1, k'_2, k'_3) \delta_K(\bk'_{123}),
\end{equation}
and $N_{123}$ is the number of triangles that contribute to a bin, which grows rapidly with wavenumber~(c.f. \cref{eqn:ntri}). In the Eulerian binning scheme, the equilateral bin with central value at $k_{\rm equi} \approx  0.151\ihmpc$ and width $\Delta k = 2k_f \approx 0.0123 \ihmpc$ contains $N_{123} \approx 8.6 \times 10^6$ triangles. While for the tree level spectrum the discrete triangle average can be evaluated, alternatives have been proposed such as approximating the sum as a continuous integral and applying `discreteness weights'~\citep{Ivanov_2022}, or evaluating the bispectrum at an `effective triangle'~\citep{Sefusatti10}
\begin{equation}
\label{eqn:keff}
    k_i^{\rm eff} = \langle |\bk_i| \rangle_{\bk_i\in {\rm bin}},
\end{equation}
where the average is taken over all triangle configurations that land in a bin. In Fig.~\ref{fig:btreelattice} we show the relative deviation between the tree-level bispectrum computed on the lattice and the analytic predictions from different averaging schemes. Evaluating the bispectrum at the center of the bin incurs large errors reaching 30\%. The effective wavenumber prescription reduces the bias in the calculation, but we find it is still well above the statistical uncertainties of the measurement. Performing the full discrete average brings the uncertainties to sub-1\%, within the statistical errors on the mean bispectra for nearly all points. While averaging over discrete triangles produces accurate results, at $\sim 10^7$ triangles in a bin it is clearly infeasible for all but tree-level predictions. We will return to this point in \S~\ref{subsubsec:analyticbk}, but for now consider that we can suitably predict the mean control variate for the tree-level Eulerian case. 

\subsubsection{Higher-order contributions with no mean}
Given the second-order density field, we can also construct the combination $B^{221}$, which is 5th-order in the density. Being comprised of an odd number of density fields, the $B^{221}$ bispectrum has zero mean. Nevertheless, its addition should improve the performance of the Eulerian control variate. In order to include it, we should include all diagrams that enter at 5th order, which includes the $B^{311}$ bispectrum. In the $\epsilon$ expansion described in Appendix~\ref{app:treebk}, these terms contribute at $O(\epsilon^2)$. An appealing aspect of including these $\epsilon^2$ terms is that, since their mean is zero, their inclusion is `free' from the perspective of building a control variate whose mean is known analytically. Since we previously established that the binned tree level bispectrum can, in principle, be calculated accurately, it is worth investigating potential gains from including the $\epsilon^2$ bispectrum. We expect these zero-mean diagrams to improve the cross-correlation because they possess non-zero covariances with the $N$-body field
\begin{align*}
    &\langle \delta^N \delta^N \delta^N | \delta^{(2)}\delta^{(2)}\delta \rangle \sim \langle \delta^N \delta^{(2)} \rangle \langle \delta^N \delta^{(2)} \rangle \langle \delta^N \delta \rangle, \\
    &\langle \delta^N \delta^N \delta^N | \delta^{(3)}\delta \delta \rangle\sim \langle \delta^N \delta^{(3)} \rangle \langle \delta^N \delta \rangle \langle \delta^N \delta \rangle,
\end{align*}
while having zero expectation value.
 \par 
To compute the $\delta^{(3)}$ contribution to the density field we return to the recursive algorithm for EPT on the grid from Ref.~\citep{Taruya_2018}, previously shown in \cref{eqn:gridspt}. In order to control for the effect of aliasing, we follow the prescription of Ref.~\citep{Taruya_2018} and apply a spherical Fourier top-hat filter with $k_{\rm cut} = 4/3 \ihmpc$ to the second order solutions, before they are used in the recursion relation to compute terms cubic in $\delta$.  Since the cubic field will only contribute in a mean-zero form to the control variate we do not have to concern ourselves with the subtleties of implementing this filtering analytically, but extensions of Eulerian control variates to the one loop bispectrum (including up to the diagram $B^{411}$) would.\par
The left panel of Fig.~\ref{fig:ecv} shows the ratio between the $N$-body bispectrum and the three forms of Eulerian bispectra we have considered so far -- Gaussian, $\mathcal{O}(\epsilon)$, and $\mathcal{O}(\epsilon^2)$. The Gaussian bispectrum clearly averages to zero, and including the $\epsilon^2$ corrections imperceptibly alters the ratio of means, as expected from analytic considerations and the assumption of a Gaussian initial field. The right panel of Fig.~\ref{fig:ecv} shows the cross-correlation coefficient between the $N$-body bispectrum and the three Eulerian control variates discussed in this section. Confirming our intuition developed for the Zeldovich approximation, we see that the inclusion of the tree-level bispectrum on top of the standard Gaussian contribution significantly increases the cross-correlation coefficient with the bispectrum measured from an $N$-body simulation. Displaying the correlation coefficients as a function of $\ksph$ also shows that the including the tree-level bispectrum \emph{tightens} the scatter in the other triangle coordinates. At $\ksph \sim 0.15 h^{-1} {\rm Mpc}$ the correlation coefficient is 80\% for the tree level control variate, 90\% for the $B^{\epsilon^2}$ and has dropped to near 40\% for the strictly Gaussian case. The inclusion of the fifth-order bispectrum extends the range of triangles over which appreciable cross-correlation is observed. \par 
Despite the improvements from pushing to higher order in Eulerian PT, there exists an unavoidable exponential decay in the cross-correlation coefficient stemming from the absence of large-scale displacements in Eulerian theory. 

\subsection{Lagrangian Control Variates -- Zeldovich Approximation and 2LPT}
\label{subsec:lagcv}
The previous sections have shown that Eulerian control variates provide a systematic form of producing correlated surrogates of $N$-body summary statistics. However, there is also the immediately clear limitation of an exponential decorrelation, as a function of $\ksph$, between the surrogate and the non-linear bispectrum. The reason for this decorrelation is tied to lack of large-scale bulk flows in EPT. The Zeldovich approximation includes these large-scale bulk flows and this drives the success of the `Zeldovich control variates' technique introduced in \cite{Kokron_2022, DeRose_2023}. We now turn to a study of the bispectrum in lattice realizations of LPT, and its suitability as a control variate compared to EPT. \par
Matter density fields in Lagrangian perturbation theory can be generated by displacing particles, located at a position $\bq$, by their Lagrangian displacement $\bPsi(\bq)$ to late-time positions $\bx = \bq + \bPsi(\bq)$. From mass conservation, the late-time density contrast field in LPT is given exactly by the relation
\begin{equation}
    1+\delta (\bx) = \int d^3q \, \delta^{D} (\bx - \bq - \bPsi(\bq)).
\end{equation}
While $N$-body simulations solve for this displacement exactly, Lagrangian Perturbation Theory provides series solutions to this displacement of the form 
\begin{equation}
    \bPsi (\bq) \approx \bPsi^{\rm ZA}(\bq) + \bPsi^{\rm 2LPT} (\bq) + \cdots,
\end{equation}
where the Fourier-space representations of the Zeldovich and 2LPT displacements are 
\begin{align}
    &\bPsi^{\tiny \rm ZA}_{\bk} =  \frac{i\bk}{k^2} \delta_{\bk}, \label{eqn:zadisp}\\
    &\bPsi^{\tiny \rm 2LPT}_{\bk} = {\int_{\bk_1, \bk_2}} \delta^D(\bk - \bk_{12}) \frac{i\bk}{k^2} \frac{3}{14} \left [ 1 - \frac{(\bk_1 \cdot \bk_2)^2}{k_1^2 k_2^2} \right] \delta_{\bk_1} \delta_{\bk_2}, \nonumber
\end{align}
and the $\bPsi^{n {\rm LPT}}$ kernels can be generated to all orders using known recursion relations, as in EPT~\citep{Matsubara_2015}. \par 

The LPT density fields are generated by displacing particles sampled from the same pre-initial condition grid and linear density field as their corresponding \quijote box. The displaced particles are deposited using cloud-in-cell deposition and the grid is corrected for this smoothing~\citep{Sefusatti_2016}. We generate three sets of Lagrangian displacements from which we compute their bispectra: 
\begin{itemize}
    \item Zeldovich displacements sampled from the initial density field smoothed by an explicit Gaussian cutoff $e^{-(k/k_{\rm cut})^2}$ with $k_{\rm cut} = 0.5 \ihmpc$. These will always be referred to as the \emph{damped ZA} sample. 
    \item Zeldovich displacements sampled from the full initial conditions without damping.
    \item 2LPT displacements sampled from the full undamped initial conditions. 
\end{itemize}
Damping the initial power spectrum before sampling displacements will be important to match the analytic calculation of observables in ZA with the lattice representation, which we discuss shortly. This damping was also required in the case of the power spectrum~\citep{Kokron_2022}. {Since we cannot analytically compute the resummed 2LPT bispectrum, we do not consider damped 2LPT displacements in this work.}
\label{sec:zcv}
\begin{figure*}
    \centering
    \includegraphics[width=\linewidth]{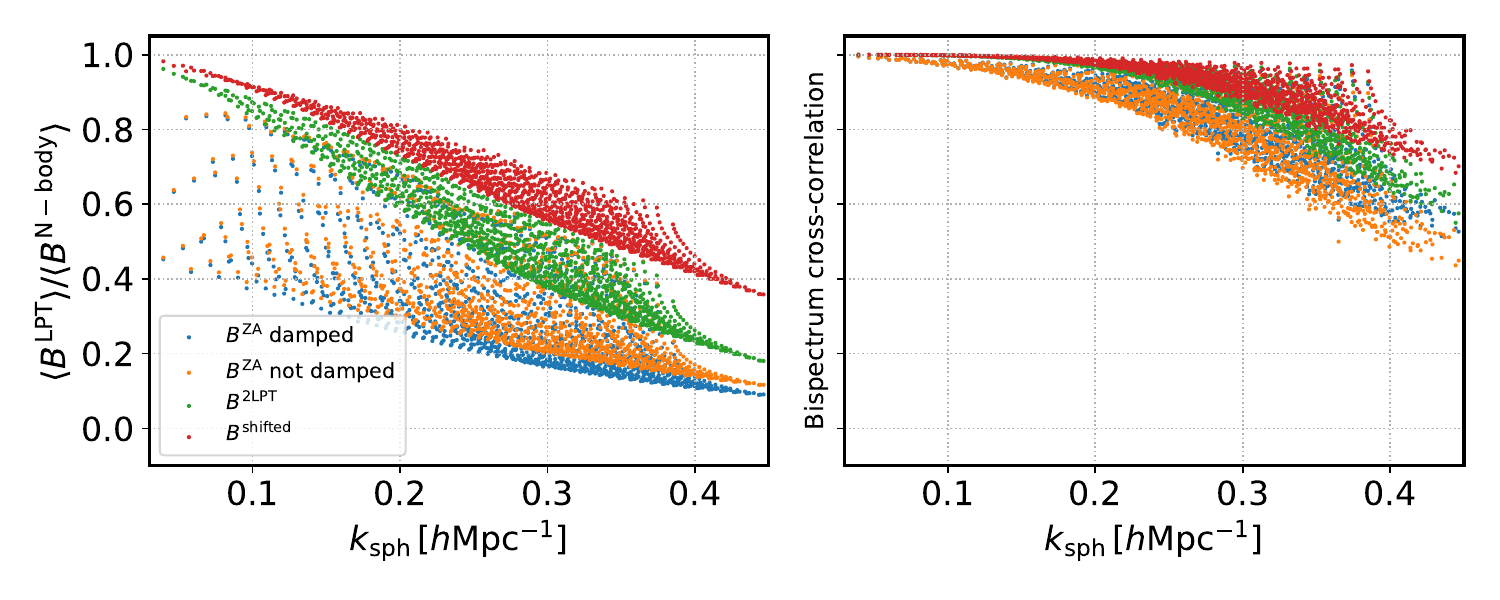}
    \caption{\emph{Left:} Ratio of bispectra in Lagrangian perturbation theory, to 2nd order in displacements, divided by the $N$-body bispectrum. Blue points correspond to the Zeldovich bispectrum, orange points to the Zeldovich bispectrum of undamped initial conditions, and the green points to the 2LPT bispectrum. Red points correspond to the bispectrum of the `shifted control variate' introduced in \S~\ref{subsec:shiftcv}. \emph{Right:} Cross-correlation coefficients between the full $N$-body bispectrum and different realizations of the bispectrum from lattice Lagrangian perturbation theory. We observe a high degree of cross-correlation across all triangles, especially when contrasted to the EPT case. However, note the Zeldovich bispectrum cross-correlation deviates slightly from $\rho = 1$ at low-$\ksph$. Despite the higher degree of correlation it is also clear the LPT bispectrum is a worse approximation to the $N$-body bispectrum than the tree-level prediction. }
    \label{fig:lptbk}
\end{figure*}

\subsubsection{Empirical LPT bispectra}
We measure LPT bispectra for the three sets of LPT displacements discussed previously. Figure~\ref{fig:lptbk} is the Lagrangian analog to Fig.~\ref{fig:ecv}: the ratio of LPT bispectra relative to the $N$-body bispectrum is shown in the left panel, while the right panel shows the corresponding cross-correlation coefficients as a function of $\ksph$. \par  
As expected from the discussion in \S~\ref{sec:ZAbk}, the tree-level bispectrum in the Zel'dovich approximation differs from the EPT prediction\footnote{{This can be seen from the structure of the second-order perturbations in each case -- \cref{eqn:za2field} for Zeldovich and \cref{eqn:ept2field} for Eulerian PT. They differ in coefficients for the `growth' and `tidal' contributions.}} As a result, even at low $\ksph$ the ZA bispectra disagree from the $N$-body bispectra, by up to 40\%. Damping the initial conditions only has a mild effect on the resulting spectra. Turning to the cross-correlation coefficient between the $N$-body bispectrum and the Zeldovich bispectrum, we observe two interesting trends. That the two bispectra disagree at very low $\ksph$ implies that the cross-correlation coefficient does not asymptote to $\rho_{x,c} \to 1$. At arbitrarily low-$k$ the Zeldovich bispectrum should not yield as much variance reduction as the EPT bispectrum. However, by virtue of including large-scale bulk flows we also observe that for the smallest-scale triangles the cross-correlation coefficient is still on the order of 50\%, whereas the Eulerian bispectra have fully decorrelated. While the damped ZA bispectrum is a worse fit to the $N$-body amplitude, we also observe that the cross-correlation coefficient at high-$k$ is slightly larger for the damped simulations. At low redshifts, past shell-crossing, the Zeldovich approximation is known to over-predict the displacement of particles compared to the $N$-body result. By smoothing the initial conditions, the displacements are slightly smaller in magnitude, and this results in a smaller overshoot and better cross-correlation with the $N$-body bispectrum. \par 
{The fact that} the damped ZA bispectrum is more correlated with the $N$-body result is interesting: it points to the possibility of engineering a filtering scale for the linear density field which maximizes the ZA correlation with the $N$-body result, without paying a price in calculating the mean. However, this optimization is probably quantitatively minor and we leave the investigation of how to determine the optimal filtering scale to future work. \par 
Turning to the 2LPT bispectrum, Fig.~\ref{fig:lptbk} shows a better agreement with the $N$-body result, especially for the lowest $k$-binned triangles, but a similarly quick disagreement as $\ksph$ increases. The 2LPT result is only 20\% of the $N$-body amplitude for the smallest scales considered. Unlike for the Zeldovich approximation, the cross-correlation coefficient between the 2LPT bispectrum and the $N$-body bispectrum asymptotes to 1 at large scales, and decays similarly but with a larger amplitude for all triangles. This is attributed to the presence of large-scale displacements, while containing the tree-level bispectrum. \par 
Quantitatively, the 2LPT bispectrum has $(1-\rho^2)^{-1} \sim 10^4$ for the lowest $\ksph$ bins. Since the full 2LPT bispectrum cannot at present be easily computed analytically this also implies, from \cref{eqn:cvN}, that at least $10^4$ 2LPT boxes at similar volume would be needed to ensure a sufficiently precise determination of the mean.
\subsubsection{The analytic Zeldovich bispectrum}
\label{subsubsec:analyticbk}
The large number of simulations needed to measure the 2LPT bispectrum to sufficient precision for it to be a suitable control variate motivate studying the Zeldovich bispectrum further. Despite not fully correlating with the matter bispectrum at low-$k$, the decorrelation is slower as a function of $\ksph$. Another key advantage of the Zeldovich bispectrum is that it is analytically calculable. Simultaneously employing an Eulerian and Zeldovich bispectrum as control variates (and extending the problem to two Lagrange multipliers $\beta_{\epsilon^2},\beta_{\rm ZA}$, for example) would allow for a set of control variates which are correlated over all scales and analytically calculable, for any triangle configuration. \par  {We turn to discuss how to compute the Zeldovich bispectrum analytically and compare it to the binned bispectrum measured from lattice Zeldovich realizations. The higher computational cost of evaluating $B_{\rm ZA}$ analytically will motivate the construction of `weights' which significantly reduce the number of evaluations of $B_{\rm ZA}$. However, for very wide bins at high $k$ these weights are also challenging to compute, and we will introduce a Monte-Carlo approximation to evaluate bin-averaging in the limit of a large number of triangles.} \par

In \cite{chen2024bispectrumlagrangianperturbationtheory} it was shown that the Zeldovich matter bispectrum can be written in a closed form 
\begin{widetext}
    \begin{equation}
    \label{eqn:analytizabk}
        B_{\rm ZA} (\bk_1, \bk_2) = \int_{\bq, \br} e^{-i \bk_1 \cdot \bq - i \bk_2 \cdot \br } \exp \left [ \frac{1}{2} (k_{1,i} k_{3,j} A_{ij} (\bq) + k_{2,i} k_{3,j} A_{ij} (\br) + k_{1,i} k_{2,j} A_{ij} (\bq - \br) ) \right ], 
    \end{equation}
\end{widetext}
where $A_{ij}(\bq)$ is the standard Zeldovich pairwise displacement correlator
\begin{equation*}
    A_{ij} (\bq) = 2\int_{\bk} (1 - e^{i\bk \cdot \bq}) \left ( \frac{k_i k_j}{k_4} \right ) P_{\rm lin}(k).
\end{equation*}
\cref{eqn:analytizabk} can be cast into a convolution of three scalar functions 
\begin{equation*}
    \mathcal{E} (\bk_i, \bk_j, \bp) \equiv \int_{\bq} e^{-i \bp \cdot \bq } e^{\frac{1}{2} \bk_{i,a} \bk_{j,b} A_{ab} (\bq)}
\end{equation*}
which can be efficiently numerically evaluated using fast Fourier transforms.\footnote{This implementation is publicly available in the python package \href{https://github.com/sfschen/triceratops}{\texttt{triceratops}.}} The scalar $\mathcal{E}(\bk_i, \bk_j, \bp)$ functions couple long- and short-wavelength modes to arbitrary order, in addition to the coupling implied by their convolution; by damping the linear spectrum, we are more immune to the effects of specific regularization schemes and can thus more easily match analytic predictions to the lattice-based realizations. \par 

The main challenge with using the analytic Zeldovich bispectrum, then, is to compute the bin-averaged bispectrum that is measured in $N$-body simulations. As discussed in the case of the tree-level Eulerian bispectrum in \S~\ref{subsec:treecv}, evaluating the bispectrum over either the central triangle or the effective triangle incurs unacceptably large errors between the empirical measurement and its analytic prediction. In the case of the tree-level bispectrum we were able to compute the average over all triangles that contributed to the bin (although note this becomes significantly more difficult for the Zeldovich binning scheme which extends to $\ksph^{\rm max} \approx 0.5 \ihmpc$). In the case of the Zeldovich bispectrum this is no longer feasible. While being substantially faster than direct integration, discrete evaluation over large numbers of $k$ modes is still prohibitively expensive for the FFT-based predictions of the Zeldovich bispectrum, which takes on the order of $\sim 1$ second on modern computers.\par 
To compute the Zeldovich bispectrum at sufficient accuracy, we devise weights to correct for binning and discreteness effects. If the Zeldovich bispectrum is similar in amplitude to a bispectrum that can be exactly bin-averaged, the effective triangle prediction can be re-weighted by the deviation of this second bispectrum. For example, assuming we assess triangles where $B_{\rm Zel} \approx B_{\rm tree}$ (note that `tree' denotes the Zeldovich tree-level bispectrum of \cref{eqn:bktreeza}), we can write the bin-average of the Zeldovich bispectrum as
\begin{align}
    \langle B_{\rm Zel} \rangle &= \left( \frac{\langle B_{\rm Zel} \rangle}{B_{\rm Zel}^{\rm eff}} \right) B_{\rm Zel}^{\rm eff} \nonumber \\
    &= \left( \frac{\langle B_{\rm tree} \rangle + \langle \Delta B \rangle}{B_{\rm tree}^{\rm eff} + \Delta B^{\rm eff}} \right) B_{\rm Zel}^{\rm eff}, \quad \Delta B = B_{\rm Zel} - B_{\rm tree} \nonumber \\
    &= \left( \frac{\langle B_{\rm tree} \rangle}{B_{\rm tree}^{\rm eff} } \right) B_{\rm Zel}^{\rm eff} \left ( 1 + \frac{ \langle \Delta B\rangle - \Delta B^{\rm eff}}{\langle B_{\rm tree} \rangle} + \ldots\right) \nonumber \\
    &= \left( \frac{\langle B_{\rm tree} \rangle}{B_{\rm tree}^{\rm eff} } \right) B_{\rm Zel}^{\rm eff} \left( 1 + \mathcal{O}\left(\Delta \ln k\right)^2 \mathcal{O}(P) \right), \nonumber
\end{align}
where we've used that $\langle B_{\rm tree}\rangle / B^{\rm eff}_{\rm tree} - 1 = \mathcal{O}(\Delta \ln k^2)$ in the third line, since the effective wavenumber approximation cancels any errors linear in the relative bin widths. \par 
The relative deviation between the bin-average and the effective triangle approximation, for the calculable tree-level case, can be used to define weights 
\begin{equation}
    w(k_1,k_2,k_3) = \langle B_{\rm tree} \rangle / B_{\rm tree}^{\rm eff}
    \label{eqn:weights}
\end{equation}
that correct for binning effects, similar to the discreteness weights employed in \citet{Ivanov_2022} except that in this case the nonlinear structure of the bispectrum is known with no free coefficients. Corrections to this weighting scheme are suppressed by the closeness of the tree-level and fully nonlinear calculations -- governed by the size of loop corrections of order $\mathcal{O}(P)$ -- along with the narrowness of the bin $\Delta \ln k$. In practice, we use undamped linear power spectra to compute these weights rather than the damped versions used in the simulations, finding that this leads to slightly better performance. This is due to the damping dominating the scale dependence on small scales, where mode coupling contributes significantly to $B_{\rm Zel}$.

\begin{figure}
    \centering
    \includegraphics[width=1.0\linewidth]{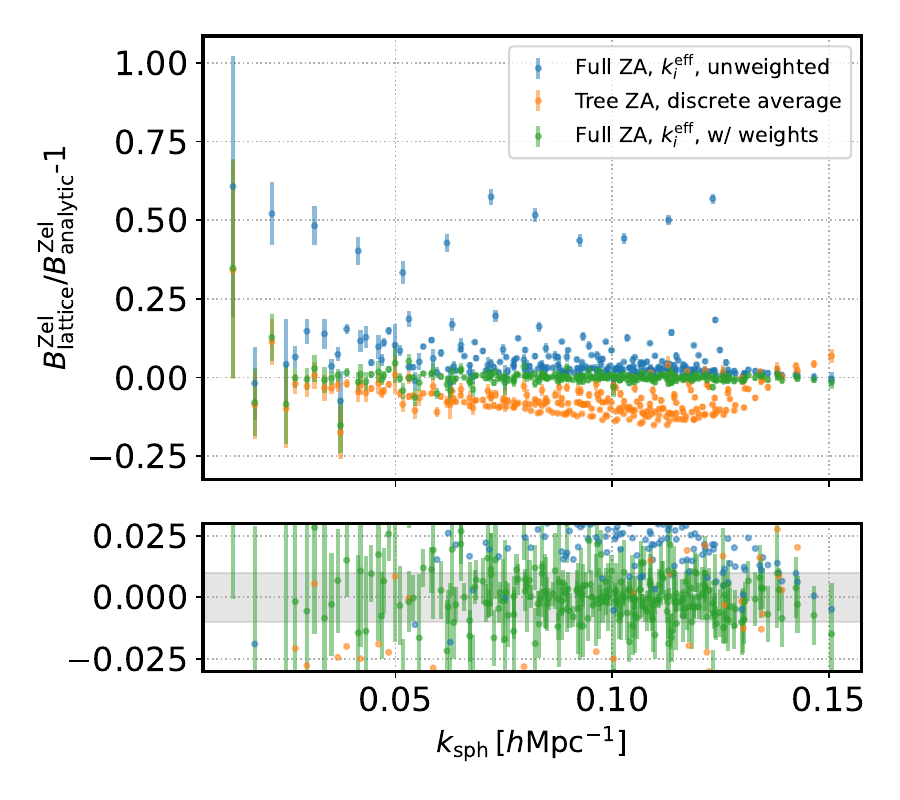}
    \caption{Analytic predictions for the Zeldovich bispectrum, including the impact of the weights in \cref{eqn:weights},
    compared to lattice-based realizations of the Zeldovich density field.  The weighted fully nonlinear predictions from \texttt{triceratops} (green) agree well with the simulations on all scales shown. The tree-level Zeldovich predictions only agree at $k \lesssim 0.05 h$ Mpc$^{-1}$ (orange). The effective triangle approximation is inadequate for the full Zeldovich bispectrum (blue), which disagrees at $\sim 10\%$ for nearly all triangles and reaches $\sim 50\%$ for squeezed triangles.}  
    \label{fig:analytic_nomc}
\end{figure}

Figure~\ref{fig:analytic_nomc} shows the bispectrum measured from lattice realizations of the Zeldovich density field compared to analytic predictions employing various approximations, using the Eulerian binning scheme. The fully nonlinear prediction computed using \texttt{triceratops} with tree-level binning weights in Equation~\ref{eqn:weights} (green) are in excellent agreement with the simulated bispectra, with a combined $\chi^2$ across all measurements -- assuming a diagonal covariance measured from 1000 simulations -- very close to $1$ per bin. In comparison, the Zeldovich tree level predictions of \cref{eqn:bktreeza}, shown in orange, agree with the fully nonlinear prediction at very large scales ($k < 0.05 h$ Mpc$^{-1}$) but rise to roughly $10\%$ by $k < 0.10 h$ Mpc$^{-1}$. The unweighted but fully nonlinear predictions shown in blue are consistently different from the green points at the $10\%$ level on all scales, reaching nearly $50\%$ discrepancies for the most squeezed triangles where the longest leg has $k_1 = 2 k_f$. These results demonstrate that the tree level weights are sufficient to restore concordance with lattice measurements without performing the costly averages over triangle configurations. \par  
As previously argued, an advantage of using Zeldovich control variates compared to Eulerian ones is their ability to remain correlated until smaller scales, or larger $k$. At these larger $k$ it is computationally more efficient to evaluate bispectra over wider bins; however, for sufficiently wide $k$ bins, even computing the tree-level prediction summing over discrete $k$ modes is computationally intensive, making it nontrivial to compute the tree-level weights in Equation~\ref{eqn:weights}. For a very wide bin, it is unlikely that any individual triangle configuration significantly impacts the results of the bin average and a Monte-Carlo sampling of valid triangles should provide a good approximation to the full average over discrete pairs. \par 
Assuming that the bispectrum is reasonably smooth within a triangle bin, the number of discrete pairs required to achieve an accuracy $\epsilon$ is roughly $N_{\rm toler} = \epsilon^{-2}$. We can therefore choose to evaluate the bispectrum a total number of $N = \text{min}(N_{\rm toler}, N_{123})$ times.\footnote{In practice we estimate $N_{ 123}$ using the continuous approximation in \cref{eqn:ntri}.} We thus downsample the number of wavevectors in bins $k_1 \pm \Delta k, k_2 \pm \Delta k$ via
\begin{equation}
    N_{1,2}^{\rm downsample} = \sqrt{f_{mc} N_1 N_2}, \quad f_{mc} = \frac{N}{N_{\rm tri}},
\end{equation}
where $N_i$ is the original numbers of wavevectors in that bin. As in the un-downsampled case, not all of the $N_1^{\rm downsample} \times N_2^{\rm downsample}$ triangles will produce a triangle that falls into the given bispectrum bin---rejecting these results in a total of $N_{\rm tri}^{\rm downsample}$ bispectrum configurations. The estimate of the binned bispectrum is the average value of the bispectrum evaluated at the remaining $\approx f_{\rm mc} N_{\rm tri}$ points. An alternative scheme is to instead downsample pairs $(\bk_1, \bk_2)$ that satisfy the geometric constraint of the bin ($|\bk_1 + \bk_2| \approx k_3$) by a factor of $f_{\rm mc}$; we have checked that this method returns a very comparable degree of accuracy but is significantly more memory-intensive to run efficiently due to having to sample the product space of wavevectors rather than the wavevectors themselves.

\begin{figure}
    \centering
    \includegraphics[width=1.0\linewidth]{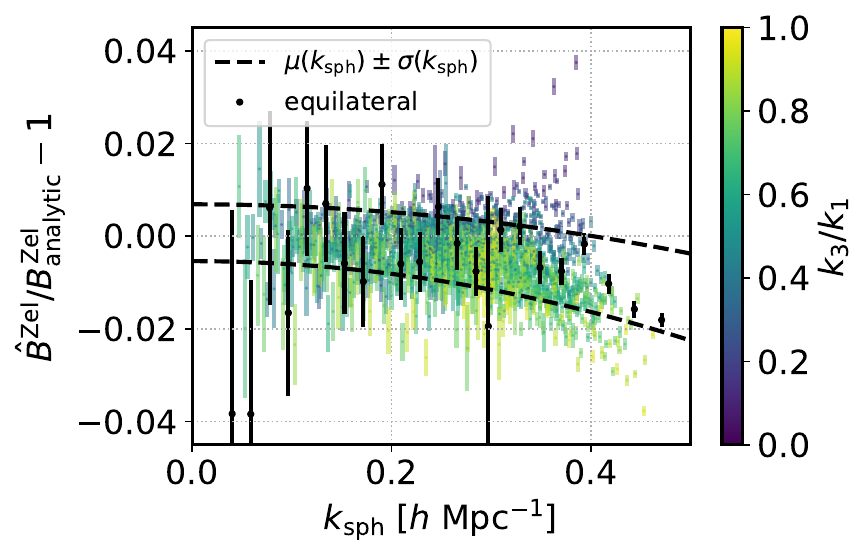}
    \caption{Similar to Figure~\ref{fig:analytic_nomc} except for the wider Zeldovich binning scheme, where the tree-level weights have to be computed by Monte-Carlo. The predictions come with a theoretical error slightly smaller than one percent, shown here as a confidence interval dependent on $k_{\rm sph}$ in the black dashed lines, as is expected given the bin width and the slope of the linear power spectrum; however, the theoretical errors are quite dependent on triangle configuration (colorbar), with the squeezed configurations most deviant while equilateral triangles (black) are statistically consistent at all but the smallest scales shown.}
    \label{fig:analytic_mc}
\end{figure}

Figure~\ref{fig:analytic_mc} shows predictions for the Zeldovich bispectrum computed using \texttt{triceratops} and tree-level weights in the wider Zeldovich binning scheme. In contrast to the finer binning scheme used in Figure~\ref{fig:analytic_nomc}, the tree-level weights in this case have been computed by Monte-Carlo, as the bins contain too many triangles to evaluate the weights. Unlike in the previous binning, the analytic predictions in this case differ from the simulations in a statistically significant way. The dashed lines in Fig.~\ref{fig:analytic_mc} show the band spanned by $1\sigma$ around a quadratic fit to the mean relative error, assuming it only depends on $\ksph$.\footnote{In particular, we maximize the log-likelihood
\begin{equation}
    \ln L = \sum_i -\frac12 \frac{ (\Delta_i - \mu_{\rm th}(k_{\rm sph, i}))^2}{\sigma^2_{\rm th}(k_{\rm sph, i}) + \sigma^2_i} - \frac12 \ln(\sigma^2_{\rm th}(k_{\rm sph, i}) + \sigma^2_i)
\end{equation}
where the relative error in each bispectrum bin is $\Delta_i = (B_{\rm pred, i} - \hat{B}_i)/\hat{B}_i$ and $\sigma_i$ is its standard deviation, assuming that $\mu_{\rm th}, \sigma_{\rm th}$ are polynomials of a given order in $k_{\rm sph}$. We estimate the theoretical and total (theoretical and statistical) uncertainty as a function of $k_{\rm sph}$ by fitting this likelihood with and without the addition of $\sigma_i^2$, finding that the theoretical error reaches about $50\%$ of the total variance at $k_{\rm sph} \approx 0.2 h$ Mpc$^{-1}$ fairly independently of polynomial order.
} The differences are consistent with a roughly $0.7\%$ theoretical error on our analytic predictions, though the theoretical error at low $k_{\rm sph} \lesssim 0.2 h$ Mpc$^{-1}$ is subdominant to statistical uncertainties.

This is inline with expectations that the error is of order $(\Delta \ln k)^2$, taking into account the slope of the linear power spectrum. This isotropic summary of the theoretical error only partially captures the story, however. The deviations are very strongly dependent on the orientation (shown here using $k_3/k_1$ as a proxy), with squeezed triangles as clear outliers in their deviation. In contrast, equilateral bins, highlighted in the black points, remain statistically consistent between the simulations and predictions until $k_{\rm sph} \approx 0.3 h$ Mpc$^{-1}$, where the effects of grid-level smoothing also becomes rather significant. 

One more numerically intensive but potentially useful solution is to construct interpolation tables of the Zeldovich bispectrum computed by \texttt{triceratops} as a function of $k_1$, $k_2/k_1$ and $k_3/k_1$. The latter two are dimensionless quantities with ranges $[0,1]$, making them more intuitive targets for interpolation. \par 
Nevertheless, we have shown that the Zeldovich bispectrum can in principle be evaluated analytically at the accuracy required for it to be a successful control variate, especially for large-scale triangles where the bispectrum variance is significant. 
\subsection{Shifted control variates}
\label{subsec:shiftcv}
Previous sections considered bispectrum control variates in Eulerian and Lagrangian theory. Eulerian control variates captured the tree-level correlation but decayed exponentially due to their lack of resummed displacements. The Zeldovich bispectrum maintained a high correlation with the $N$-body bispectrum but lacked the correct low-$k$ form and would not optimally cancel variance if used as a control variate. Is there a way to engineer a control variate which only resums linear displacements, but possesses the correct tree-level bispectrum (or $N$-point function more generally)?\par 
The `shifted operator' scheme for perturbation theory of \cite{Schmittfull_2019} can be used to this end. Shifted operators keep only the Zeldovich displacement exponentiated and in Fourier-space, the shifted operator of a field $\mathcal{O}(\bq)$ is defined as
\begin{equation*}
    \tilde{\mathcal{O}}(\bk) = \int_{\bq} e^{-i \bk \cdot (\bq + \bPsi^{\rm ZA}(\bq))}\mathcal{O}(\bq).  
\end{equation*}
Contributions from higher-order displacements can be expanded from the exponential, as they are small compared to the Zeldovich displacement. The tree-level bispectrum contains the quadratic bias fields $\delta^2$ and $s^2$. Consider then, instead of the Eulerian $\delta^{(2)}$ field, using the bispectrum measured from a Lagrangian `biased tracer'
\begin{figure}
    \centering
    \includegraphics[width=\linewidth]{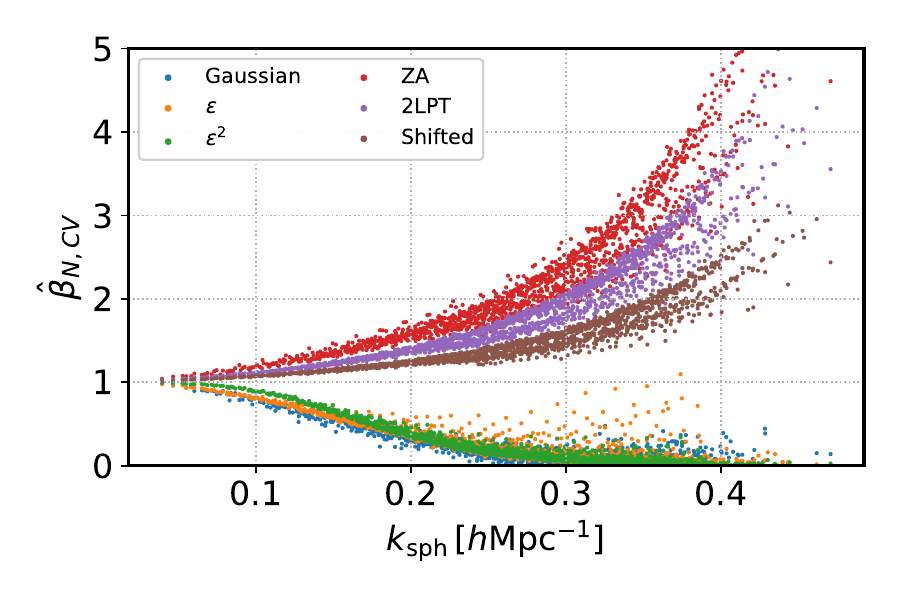}
    \caption{Estimates of the Lagrange multiplier, $\beta$, from the estimators of \cref{eqn:univariatebeta} for the various control variates considered in this work. The solid lines show the median and 95\% quantiles along bins of $\ksph$. }
    \label{fig:betaest}
\end{figure}

\begin{widetext}
\begin{equation}
\label{eqn:lptcv2}
    \delta^{\rm CV, (2)} (\bk) = \int d^3q e^{-i\bk \cdot \bq} e^{-i \bk \cdot \bPsi^{\rm ZA}(\bq)} \left [ 1 + c_1 \delta(\bq) + c_2 \delta^{2}(\bq) + c_s s^2 (\bq) \right ].
\end{equation}
Expanding linearly in the Zeldovich displacement, the Eulerian version of this field is 
\begin{equation}
    \delta^{\rm CV, (2)} (\bx) \approx (1+c_1)\delta (\bx) + \left( \frac{2}{3} + c_1 + c_2 \right ) \delta^2 (\bx) - (1+c_1)\bPsi \cdot \nabla \delta + \left ( \frac{1}{2} + c_s \right) s^2 (\bx) + O(\delta^3; c_1, c_2,c_s),
\end{equation}
where the $2/3$ and $1/2$ terms come from the Eulerian expansion of the Zel'dovich kernel, \cref{eqn:za2field}. Setting $\{c_1, c_2, c_s\} = \{0, 1/7, -3/14\}$ results in a surrogate density field which is in agreement with the second order Eulerian PT solution. Its bispectrum at tree level is given by the Eulerian tree level bispectrum\footnote{Note that we could also re-write the `1+' term in \cref{eqn:lptcv2} a sum of shifted operators with well-defined coefficients, which will change the value of the $\{c_i\}$ that recover the correct tree-level expression. The resulting expression will be different at 3rd order compared to using the `1+' term. }. However, the field is fundamentally Lagrangian and so it will not pay the price of exponential decorrelation. \par
An advantage of constructing an operator in this fashion is that since only Zeldovich displacements are exponentiated, their bispectra can in principle be calculated analytically to all orders. The expressions will involve integrals similar to \cref{eqn:analytizabk}\footnote{The expression in \cref{eqn:analytizabk} is equivalent to \cref{eqn:biasbk} if we set all $\mathcal{O}_i(\bq) = 1$.}
\begin{equation}
\label{eqn:biasbk}
    \int_{\bq, \boldsymbol{r}} e^{- i \bk_1 \cdot \bq - i \bk_2 \cdot \boldsymbol{r}} \left \langle \mathcal{O}_i (\bq_1) \mathcal{O}_j (\bq_2) \mathcal{O}_k (\bq_3) e^{-i \bk_1 \cdot \bPsi (\bq_1) - \bk_2 \cdot \bPsi (\bq_2) + i (\bk_1 + \bk_2) \cdot \bPsi (\bq_3)} \right \rangle, 
\end{equation}
\end{widetext}
which can be evaluated using the cumulant expansion theorem. The resulting expressions for the basis fields of the LPT bispectrum can be found in Appendix E of \cite{chen2024bispectrumlagrangianperturbationtheory}. While explicitly written down, the efficient numerical implementation of these bispectra is still an open task. {This is because of the additional angular structure imposed by anisotropic mode coupling on the bispectrum, e.g. because of the tidal coupling $s_{0,ij} s_{0,ij}$, which require many new angular kernels beyond the Zeldovich matter field in the formalism employed in \texttt{triceratops}.} We will thus explore the applicability of this hybrid scheme, dubbed `shifted control variates', by numerically estimating their bispectra. A calculation of the mean of the control variate to high accuracy is only needed when computing the unbiased average. An assessment of the total amount of variance reduction achieved by a control variate does not require this. Thus, we will study the shifted control variate to motivate future development of calculations of the analytic Zeldovich bispectrum for biased tracers. The left and right panels of \cref{fig:lptbk} show the shifted control variate compared to the other Lagrangian models assessed in this work. We clearly see that not only is the shifted control variate the closest approximation to the $N$-body bispectrum, it is also the most correlated and decays at a slower rate than even the 2LPT bispectrum. For the highest $\ksph$ triangles the shifted control variate is found to be $\sim$70\% correlated while the 2LPT coefficient is near $\sim$60\%. 
\section{Results and Discussion}
Having characterized our ability to analytically estimate the binned perturbative bispectrum, as well as the correlation coefficient of several Eulerian and Lagrangian control variate candidates, we turn to the study of the variance reduction offered by each candidate. 
\label{sec:results}
\subsection{Estimates of $\beta$}
We estimate the Lagrange multiplier, $\beta$, for each control variate from our set of $N_{\rm sims}=1000$ simulations. {\cite{chartier2020} presented different estimates for the Lagrange multiplier which solved either a univariate (bin-by-bin) or multivariate optimization problem. In this work wee estimate the Lagrange multiplier only in the univariate approximation}
\begin{equation}
\label{eqn:univariatebeta}
    \hat{\beta}_{N,{\rm CV}}(k_1, k_2, k_3) = \frac{{\rm diag}({\rm Cov}(\hat{B}_N, \hat{B}_{\rm CV}))}{{\rm diag}({\rm Var} (\hat{B}_{\rm CV}))},
\end{equation}
for the full set of triangles. We compare the performance of the univariate estimator with the full multivariate estimator
\begin{equation}
\label{eqn:multivariatebeta}
    \hat{\beta}_{N,{\rm CV}} (k_1, k_2, k_3) = (\Sigma_{N,{\rm CV}} \cdot \Sigma^+_{\rm CV}),
\end{equation}
in Appendix~\ref{app:betaest}, as well as the diagonal approximation of \cite{Kokron_2022}. In this appendix we also show that the observed `damping' of $\beta$ discussed in \cite{Kokron_2022} was spuriously driven by the under-determination of the covariance matrix in the univariate problem, and usage of the pseudoinverse to invert the control variate covariance.

We report the measurements of the Lagrange multiplier, for the different classes of control variates considered, in Fig.~\ref{fig:betaest}, as a function of $\ksph$.  For the Eulerian control variates we observe a monotonic and steep decrease for $\beta$ as a function of $\ksph$. This can be understood from the fact that any cross-covariance between an Eulerian control variate and an $N$-body summary statistic has to decay as $\sim \exp(-\Sigma^2 k_{123}^2)$ (as discussed in \S~\ref{sec:zerocv}). Since the Lagrange multiplier can be similarly expressed as 
\begin{equation}
    \beta = \rho_{x,c} \frac{\sigma_x}{\sigma_c}, 
\end{equation}
for an Eulerian control variate we expect exponential decorrelation. For the Lagrangian control variates shown in the right panel of Fig.~\ref{fig:lptbk}, the cross-correlation coefficient has not fallen off as aggressively, reaching $\rho \sim 0.5$ for the smallest-scale triangles considered. The left panel of Fig.~\ref{fig:lptbk} shows that $B^{\rm N-body} \approx 10 B^{\rm ZA}$ at those scales, and assuming that the bispectrum covariance at these scales is dominated by the $BB$ term\footnote{This is a very approximate scaling, but in Fig.~\ref{fig:bsnr} we see that for the highest-$k$ triangles the disconnected `PPP' contribution under-estimates the covariance by a significant amount.} 
\begin{equation*}
    \frac{\sigma(B^{\rm N-body})}{\sigma(B^{\rm ZA})} \sim \frac{B^{\rm N-body}}{B^{\rm ZA}} \approx 10,
\end{equation*}
then we expect $\beta \sim 5$ which is close to the observed value for the ZA spectrum. In concordance with the rough scalings presented here, we also expect to find a lower value of $\beta$ for the 2LPT bispectrum (since the degree of correlation is similar but $B^{\rm 2LPT} \sim 2 B^{\rm ZA}$), and we see this in Fig.~\ref{fig:betaest}. The shifted control variate's bispectrum has a Lagrange multiplier closest to unity for all triangles, compared to other methods. 
\par

\subsection{Variance reduction}
\begin{figure}
    \centering
    \includegraphics[width=1\linewidth]{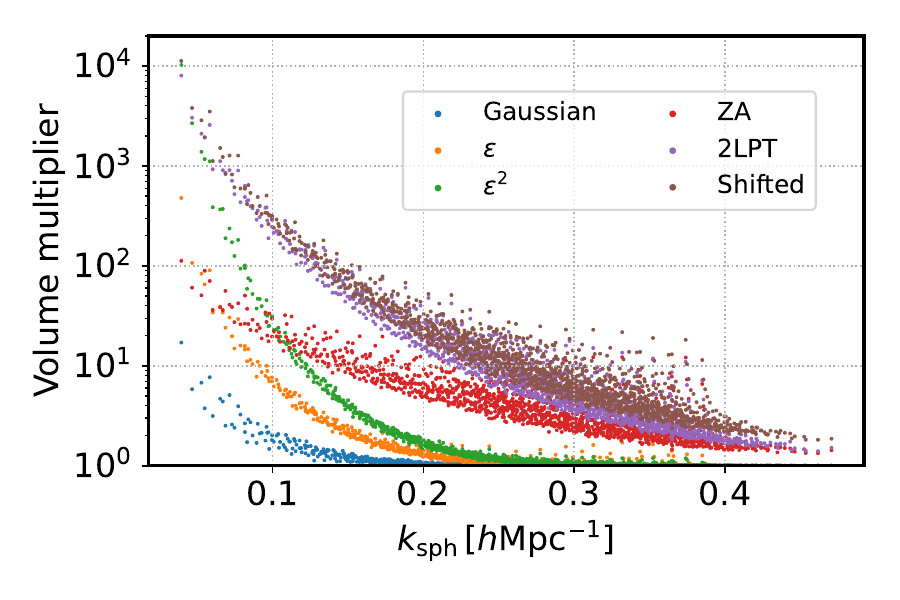}
    \caption{ Effective `volume multiplier' from applying the various bispectrum control variates considered in this work, after combining with the optimal $\beta$. The shifted operator and 2LPT control variates perform the best across most triangle configurations.}
    \label{fig:vmult}
\end{figure}

With $\beta$ in hand, we construct variance-reduced estimates of the bispectrum for each of the control variates considered in this work, for the $N=1000$ simulations in the \quijote ensemble. We measure the resulting variance from this ensemble, and compare it to the variance of the $N$-body bispectrum. In the Fig.~\ref{fig:vmult} we report the `volume multiplier' 
\begin{equation*}
    {\rm Volume\, multiplier} (k_1, k_2, k_3) \equiv \left ( \frac{\sigma^2_{N}}{\sigma^2_{\rm CV}}\right ) (k_1,k_2,k_3),
\end{equation*}
which corresponds to the effective number of simulations needed to be averaged over to achieve the same variance. {We show six sets of points, representing the full set of Eulerian and Lagrangian control variates considered in this work: 
\begin{enumerate}
    \item Gaussian control variates, as described in \S~\ref{sec:zerocv}. 
    \item Eulerian control variates to tree level ($\mathcal{O}(\epsilon)$) and including zero-mean diagrams at $\mathcal{O}(\epsilon^2)$, introduced in \S~\ref{subsec:treecv}. 
    \item Control variates in Lagrangian Perturbation Theory, including using the Zeldovich approximation and the 2LPT density field (\S~\ref{subsec:lagcv}). 
    \item The `shifted control variate' (\cref{eqn:lptcv2}), which resums only the Zeldovich displacement but can be tuned to correctly capture any $n$-point function at tree level. 
\end{enumerate}
We additionally note that, since we are only interested in verifying the amount of variance reduction achieved, the results in this section do not depend on the form used for the mean of the control variate. }
For all triangle configurations with $\ksph \lesssim 0.25 \ihmpc$, the shifted operator and 2LPT control variates achieve a 10-fold increase in volume, nearing a $10^4$-fold increase for the largest-scale triangle configurations we have considered. For the Eulerian control variates, the $\mathcal{O}(\epsilon^2)$ control variate achieves comparable volume increase to the shifted operator at the largest scales, but this increase decays and by $\ksph=0.2\ihmpc$ the improvement is negligible. \par
\begin{figure}
    \centering
    \includegraphics[width=1\linewidth]{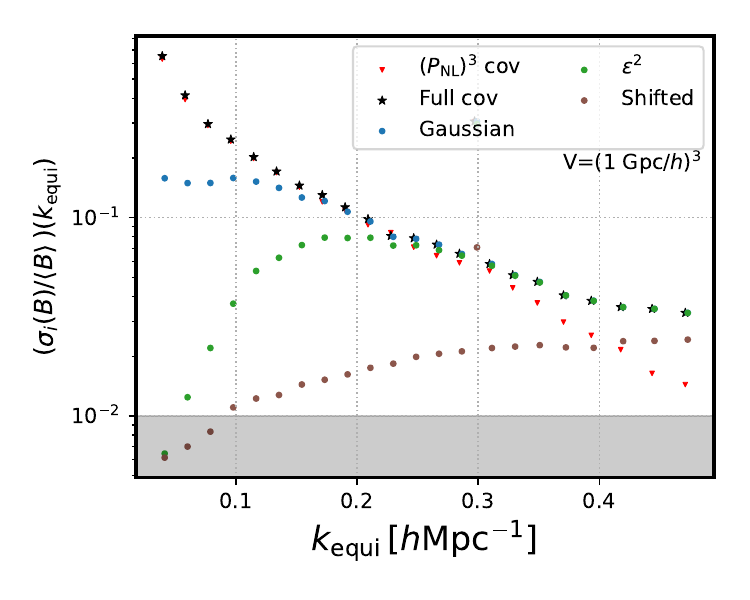}
    \caption{ Effective precision of the bispectrum for equilateral triangles, for a 1 (Gpc$/h)^3$ box, after applying the Gaussian, best-performing Eulerian, and best-performing Lagrangian control variates. The red triangles denote the bispectrum precision inferred when assuming the `PPP' disconnected term of the covariance only, and the black stars show the precision inferred from using the full numerical covariance, without applying any control variate. }
    \label{fig:bsnr}
\end{figure}

 
We present another view of the results of applying different control variates to the bispectrum in Fig.~\ref{fig:bsnr}\footnote{{There is a numerical artifact in the measurement for the point slightly below $k_{\rm equi}=0.3\ihmpc$, caused by the heterogeneous binning scheme we have adopted. No other triangle is affected.}}. Focusing on equilateral triangles (although the trends observed hold for all triangles), we show the precision with which the bispectrum is measured for several different forms of control variate. We also show, for comparison, the `true' precision obtained using the full nonlinear covariance and the precision inferred by assuming an approximate form for the covariance. The approximation we make is to use the `nonlinear disconnected covariance'. Specifically, for these equilateral triangle configurations we measure the bispectrum covariance from the Gaussian simulations with matched initial conditions. The covariance in this Gaussian disconnected case is 
\begin{equation*}
    \hat{\sigma}^2_{\rm gauss} (B) \sim \frac{1}{N_k} P_{\rm lin}^3(k_{\rm equi}),
\end{equation*}
where $N_k$ is the number of $k$-modes that enter that equilateral triangle bin. We compute the nonlinear disconnected covariance (also called the `PPP' covariance) by rescaling the empirical estimate of the Gaussian bispectrum covariance with the ratio of power spectra
\begin{equation}
\label{eqn:gcovnonlin}
     \hat{\sigma}^2_{\rm gauss} (B) \to  \hat{\sigma}^2_{\rm gauss} (B) \times \left ( \frac{P_{\rm nonlin}}{P_{\rm lin}} \right )^3(k_{\rm equi}).
\end{equation}
Any difference between the full covariance and \cref{eqn:gcovnonlin}, then, will arise from the importance of terms such as the bispectrum-bispectrum, power spectrum-trispectrum, and pentaspectrum terms in the covariance. This comparison reveals an interesting feature of numerical bispectrum estimation: even for the relatively broad bins considered in this work, the precision to which the bispectrum is measured at scales of $k_{\rm equi} \gtrsim 0.4 \ihmpc$ saturates at a level of 4\% for our volume of $V=1 ({\rm Gpc}/h)^3$. Using the nonlinear `PPP' covariance would lead to $O(1)$ error on the precision with which the bispectrum is measured. \par 
Turning to the performance of the different control variates we have considered, we find that employing the Gaussian control variate does not significantly increase the precision of the measurement -- there is some improvement for bins with $k_{\rm equi} \leq 0.1 \ihmpc$ but the measurement uncertainties are still around 15\%. The best performing Eulerian control variate, the bispectrum to $\mathcal{O}(\epsilon^2)$, achieves below 1\% for the first equilateral bin, but rapidly decorrelates. By $k_{\rm equi} \sim 0.2 \ihmpc$ the measurement uncertainties are comparable to the $N$-body values. Finally, we see that the shifted control variate leads to sub-1\% measurements of the bispectrum for the longest triangles, and better precision than what can be achieved at the smallest scales. As the shifted bispectrum decorrelates from the $N$-body result we approach the $N$-body precision but there is still an improvement even for the highest-$k$ triangle. The shifted control variate can essentially eliminate large-scale sample variance as a concern for bispectrum estimation, by achieving sub 2-\% precision across all triangle bins in a single $V = 1 ({\rm Gpc}/h)^3$ box. \par 
Simulation-based modeling of the bispectrum{, in principle,} is now gated by the precision to which small-scale triangle configurations can be measured. 
\subsection{Comparison to past results}
Having assessed the performance of our different perturbative control variates, we compare them to past applications of variance reduction in the published literature. \par 
The first paper to consider the performance of a control variate in the bispectrum was \cite{chartier2020}. Two sets of triangles are considered -- `squeezed isoceles triangles' with $k_1 = k_2$ from $k_3/k_1 \in [0.025, 0.2]$, and a set of equilateral triangles with $k_{\rm max} = 0.75 \ihmpc$. Their analysis uses as a surrogate a cheaper $N$-body simulation from COLA~\citep{tassev2012}, and they average their control variate estimators over five $N$-body -- COLA pairs to report their variance reduction. The effective volume multiplier they report for isoceles triangles is between 40$\times$ and $\sim 5 \times$ in this scenario, when considering a single pair of simulations. Since this squeezed isoceles analysis bins all triangles together, it is not clear how to map these volume reductions, but we note that a volume multiplier of 5 is achieved for $\ksph \sim 0.3 \ihmpc$ for our shifted control variate. Their equilateral numbers are more translatable -- focusing on their equilateral triangle with $k_{\rm min} \sim 0.04\ihmpc$ they report a per-pair volume multiplier of 200$\times$, and their equilateral triangle with $\ksph \approx 0.5 \ihmpc$ has a per-pair volume multiplier of 1.4. The shifted control variate has a volume multiplier of $\sim 1.8$ for the $\ksph = 0.47 \ihmpc$ equilateral triangle of our Lagrangian binning scheme. Thus, we find performance comparable to the original investigation of \cite{chartier2020} with a control variate that is analytically tractable\par 
The other work which has investigated control variates for the bispectrum is \cite{Ding:2022ydj}, which investigates the use of the FastPM~\citep{Feng2016} as a control variate. Additionally, \cite{Ding:2022ydj} investigate the case of the halo bispectrum which is not immediately comparable to the analysis we have carried out in this work. Nevertheless, we note that their analysis corresponds to triangles with $k_1 = 0.1 \ihmpc, k_2 = 0.2\ihmpc$ and an angle of $\theta = [0, \pi]$, which are triangles with $\ksph \in [0.14, 0.21]\ihmpc$. At these scales they note no improvement from pairing and fixing, and an improvement in volume that corresponds to a per-pair volume multiplier of around 20$\times$, under the assumption of no uncertainty in $\mu_c$. It is also interesting that their volume multiplier is somewhat flat across the triangles they consider. Across the equivalent range of scales we find the volume multiplier sharply varies from $100\times$ at $\ksph \sim 0.14 \ihmpc$ to $20\times$ at $\ksph \sim 0.21 \ihmpc$. It could be that the presence of shot noise in the case of the halo bispectrum has set an upper bound to the volume multiplier that can be obtained, but we leave an explicit comparison to halo bispectra for future work. \par 
While no past work has investigated variance reduction for the bispectrum as we have here, for cases where a comparison is possible we find the shifted control variate performs as well as approximate $N$-body solvers investigated in past works at a fraction of the computational cost. 

\section{Conclusions}
\label{sec:conclusions}
We have investigated the problem of reduction of variance in the empirical, simulation-based bispectrum using control variates inspired by Eulerian and Lagrangian perturbation theory. We used the Zeldovich approximation as a toy model for the full nonlinear problem, where the exact solution is known and the corresponding `Eulerian perturbation theory' is highly simplified, to investigate the structure of exponential decorrelation for any Eulerian control variate as well as how perturbative corrections restore some of the correlation structure. Eulerian control variates always decorrelate as $\sim \exp [ -\Sigma^2 k_{123}^2]$, with second and third order corrections serving to eliminate the leading-order suppression of the cross-correlation coefficient. Additionally, we showed that bispectra of mean-zero can still be useful control variates, suggesting that including diagrams which are naively zero in the Eulerian case can be beneficial since there is a non-zero covariance between the $N$-body and Gaussian case. \par 
We then turned to a study of Eulerian and Lagrangian control variates in the non-linear problem, where our Eulerian intuition developed in the preceding toy model holds. We find the Zeldovich approximation is not an optimal control variate in the case of the bispectrum -- unlike the power spectrum -- and show this is due to the differing tree-level structure of the bispectrum in ZA and in EPT. Still, the Zeldovich bispectrum possesses the advantage of being significantly correlated with the $N$-body case out to small scales, as in the case of the power spectrum. The 2LPT bispectrum is shown to be more optimally correlated at the cost of not being analytically calculable. We introduce the `shifted control variate' as an optimal solution -- being in principle analytically calculable while simultaneously possessing maximal correlation with the $N$-body case. Indeed, the shifted control variate is shown to outperform all other perturbative control variates in this work. The use of a single $N$-body / shifted control variate pair is shown to reduce the variance of $N$-body simulations by factors ranging from $10^4$ to $1.8$ for triangles between $\ksph = [0.036, 0.48]\ihmpc$. Equivalently, with a single $V=1({\rm Gpc}/h)^3$ box we can measure the matter bispectrum at $z=0.5$ to sub-2\% precision for all triangle configurations in question. This implies that accurate simulation-based bispectrum emulators can be devised, extending the conclusions of past work on the power spectrum to this domain. \par 
There exist clear directions to continue this work. We have focused on the case of the matter bispectrum, the scales of $k\lesssim 0.5 \ihmpc$ are currently being probed by galaxy surveys and it is of great interest to extend this technique to the bispectrum of biased tracers. We note that the methodology laid out here is fully sufficient -- the biased tracer bispectrum is given by \cref{eqn:biasbk} and is a mild generalization of the shifted operator control variate. The remaining work would be to select samples representative of the clustering samples of upcoming surveys such as DESI, Euclid and Rubin to study their bispectrum signatures, as has been investigated (using non-perturbative control variates and HODs) in \cite{ding2025suppressingsamplevariancedesilike}. {In \cite{Kokron_2022} we showed that a simple linear bias control variate can significantly reduce the variance of galaxy power spectra. For bispectra, we expect that the `tree-level' expansion using the quadratic and tidal bias operators should be sufficient to achieve comparable reductions in variance. Since the leading decorrelation in the case of the galaxy density field is shot noise, the tracer number density will is expected toset the fundamental limit on the achievable volume multiplier. }\par Another direction would be to extend this work to redshift space. The extension of Zeldovich control variates to redshift space is known for the power spectrum~\citep{DeRose_2023}. As \texttt{PolyBin3D} possesses the functionality to measure bispectrum multipoles, extending the empirical results of this paper to redshift space is readily achievable. The biggest challenge comes in estimating the mean: while the Zeldovich matter bispectrum can be computed in an identical way to the real space one, as shown in \citet{chen2024bispectrumlagrangianperturbationtheory}, in order to capture nonlinearities beyond the Zeldovich approximation one would either have to formulate redshift space lattice-based EPT and match its bispectrum to high accuracy (as in \cite{taruyafield}), or extend the LPT bispectrum to redshift space including quadratic nonlinearities in order to produce analytic predictions of the mean of the redshift space bispectrum control variate.  \par 
Finally, the construction of the shifted control variate points to the possibility of engineering a control variate whose $n$-point is correct at tree level. The extension of shifted control variates to the trispectrum could have implications in the problem of covariance matrix estimation for galaxy surveys. We plan to return to this topic, and the others mentioned above, in future work. 

\section*{Acknowledgments}
The authors thank Delon Shen and Matias Zaldarriaga for helpful discussions, and Alexander Dittmann for assistance with the OJA template. NK acknowledges support from the Bershadsky Fund and the Fund for Natural Sciences of the Institute for Advanced Study. SC acknowledges support from the National
Science Foundation at the IAS through NSF/PHY 2207583. Support for this work was provided by NASA through the NASA Hubble Fellowship grant
\#HST-HF2-51572.001 awarded by the Space Telescope Science Institute, which is operated by the
Association of Universities for Research in Astronomy, Inc., for NASA, under contract
NAS5-26555. This work was performed in part at the Aspen Center for Physics, which is supported by National Science Foundation grant PHY-2210452 and a grant from the Simons Foundation (1161654, Troyer). SC thanks the Galileo Galilei Institute for Theoretical Physics for the hospitality and the INFN for partial support during the completion of this work. Calculations and figures in this work have been made using \texttt{nbodykit} \citep{Hand_2018} and the SciPy Stack \citep{2020NumPy-Array,2020SciPy-NMeth,4160265}. This research has made use of NASA's Astrophysics Data System and the arXiv preprint server.
\begin{appendix}

\section{Cross Correlation of Nonlinear and Eulerian Fields: Perturbation Theory and IR-enhanced Diagrams}
\label{app:ptcorr}

In this appendix we will explore the cross correlation of the nonlinear Zeldovich and matter density fields with the predictions of Eulerian perturbation theory (EPT) on the lattice, in particular to understand the correlation coefficient as a function of perturbative order and the role of long-wavelength (IR) displacements.

Let us begin with the cross correlation of nonlinear and Eulerian fields in the Zeldovich approximation. We want to consider the case where an arbitrarily nonlinear vertex of the fully nonlinear Zeldovich field, $Z_{m+2n}$, is cross-correlated to a lower order Eulerian vertex $Z_{m}$. In this case all the unpaired vertices must be paired amongst themselves, leading to an integral
\begin{align}
     \int_{\bp} Z_{m+2n}(\bk_1, ... \bk_m, \bp_1, -\bp_1, ..., & \bp_n, -\bp_n )  P_{\rm lin}(p_1) ... P_{\rm lin}(p_n) \nonumber \\
     &\sim Z_{m}(\bk_1, ... \bk_m) \left( \int_{\bp} \frac{(\bk \cdot \bp)^2}{p^4} P_{\rm lin}(p) \right)^n \supset e^{-\frac12 k^2 \Sigma^2} Z_m(\bk_1, ..., \bk_m)
     \label{eqn:zn_recursion}
\end{align}
where the final expression follows from summing up all $n$ and accounting for combinatorial factors for the number of pairing possibilities out of the original $m+2n$ momenta. This implies that the correlation of the nonlinear and Eulerian Zeldovich fields at each order in the latter can be computed exactly by enumerating diagrams excluding bubbles $(k^2 \Sigma^2)$ in the former, then resumming the bubbles as an exponential. We therefore have that, for example, $\langle \delta^N \delta^{(1)} \rangle' = P_{\rm lin}(k) e^{-\frac12 k^2 \Sigma^2}$ and $\langle \delta^N \delta^{(2)} \rangle' = P_{22}(k) e^{-\frac12 k^2 \Sigma^2}$, as also derived in Equation~\ref{eqn:nz2}.

\begin{figure}
    \centering
    \includegraphics[width=0.95\linewidth]{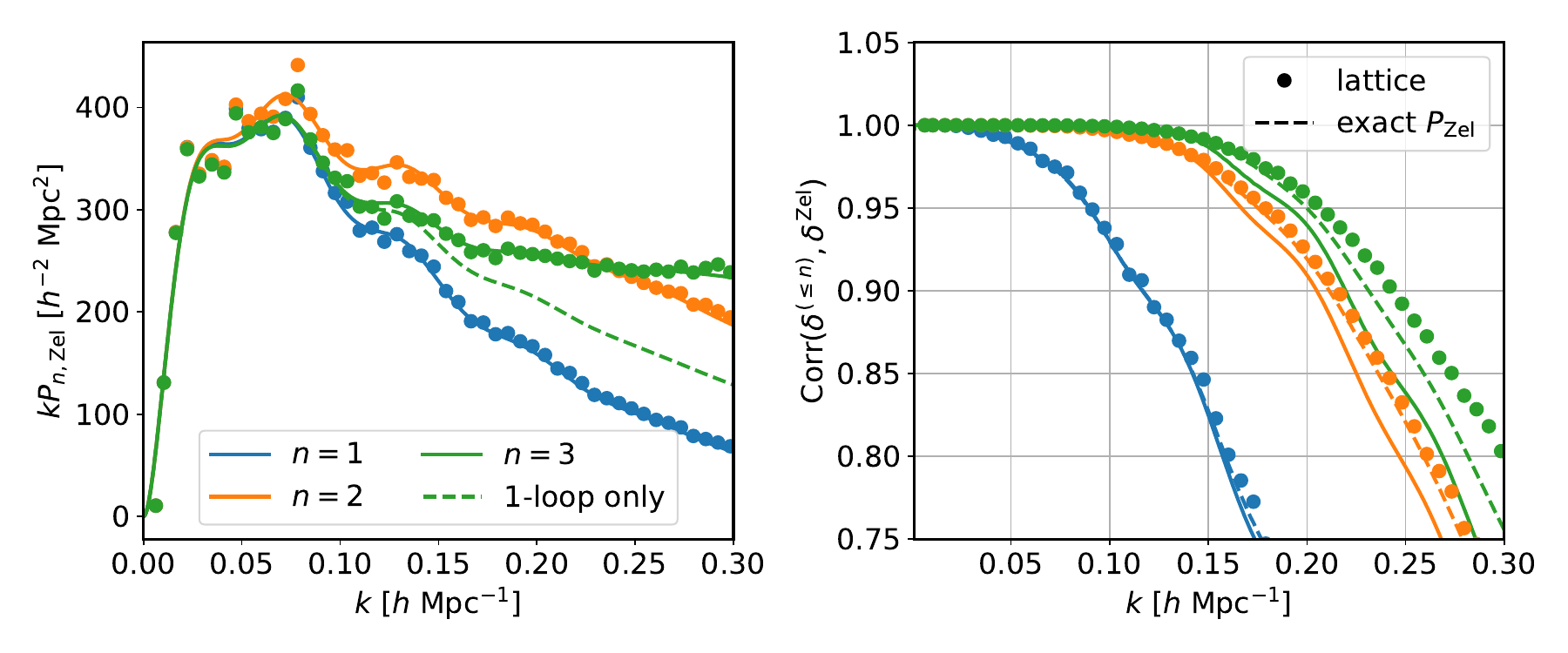}
    \caption{Cross-correlation of the nonlinear and Eulerian Zeldovich density fields on the lattice (dots) compared to analytic predictions. (Left) The cross power spectrum between the Eulerian field summed to n$^{\rm th}$ order, computed to 1-loop in perturbation theory, with large-IR contributions at higher loops resummed via an exponential. For the third-order cross correlation there is also an IR-enhanced 2-loop diagram, and we show the prediction without it as a dashed line. (Right) The correlation coefficient between the two fields. The analytic predictions break down around $k \sim 0.15 h$ Mpc$^{-1}$ reflecting the onset of 2-loop mode coupling. For $n=1,2$ the resummed expressions at 1-loop are exact except for the autospectrum of the nonlinear Zeldovich field in the operator, such that swapping out the 1-loop expression for a fully nonlinear one (dashed) is enough to achieve good agreement to smalls scales.  }
    \label{fig:zel_corr}
\end{figure}

In order to accurately predict cross correlations of this sort, as well as the autocorrelations of the Eulerian fields, it will be important to account for the large parameter $k \Sigma \gtrsim 1 $ beyond the loop order considered. This is, for example, why we have isolated out the exponential in the paragraph above, since expanding its argument to a given order would lead to known higher-order corrections that are significantly larger than naive expectations. Many of these terms are already included in the full 1-loop calculation, e.g. $P_{22} \supset (k^2 \Sigma^2) P_{\rm lin}$ and $P_{13} \supset -\frac12 (k^2 \Sigma^2) P_{\rm lin}$, where they in addition cancel when properly combined. However, when considering also the  cubic lattice EPT prediction, we also generate a subset of 2-loop diagrams whose IR contributions do not cancel, and are thus artificially enhanced. This comes from $P_{33}$, which is given by
\begin{equation}
    P_{33}^{\rm Zel, IR}(k) = \frac34 (k^2 \Sigma^2)^2 P_{\rm lin}(k) + (k^2 \Sigma^2) P_{\rm 1-loop}^{\rm Zel}(k)
\end{equation}
while the contribution to the cross correlation gives the total spectrum
\begin{equation}
    P_{3,\rm Zel}^{\rm IR-enhanced}(k) = \left( P_{13}^{\rm Zel}(k) + \frac12 (k^2 \Sigma^2)^2 P_{\rm lin}(k) + (k^2 \Sigma^2) P_{\rm 1-loop}^{\rm Zel}(k) \right) e^{-\frac12 k^2 \Sigma^2}
    \label{eqn:p3zel}
\end{equation}
where we note that the latter has a $13$ contribution due to the linear term in the Zeldovich field. In fact, the 2-loop terms in these two expressions are equal, since $P_{13}^{\rm Zel} = - \frac12 k^2 \Sigma^2 P_{\rm lin}$. The left panel in Figure~\ref{fig:zel_corr} shows this cross-spectrum compared to the measured cross-spectra in simulations. These are in excellent agreement, though for the third-order Eulerian field the IR-enhanced 2-loop diagram is critical in establishing this agreement, which we emphasize is achieved without any free parameters.

\begin{figure}
    \centering
    \includegraphics[width=0.95\linewidth]{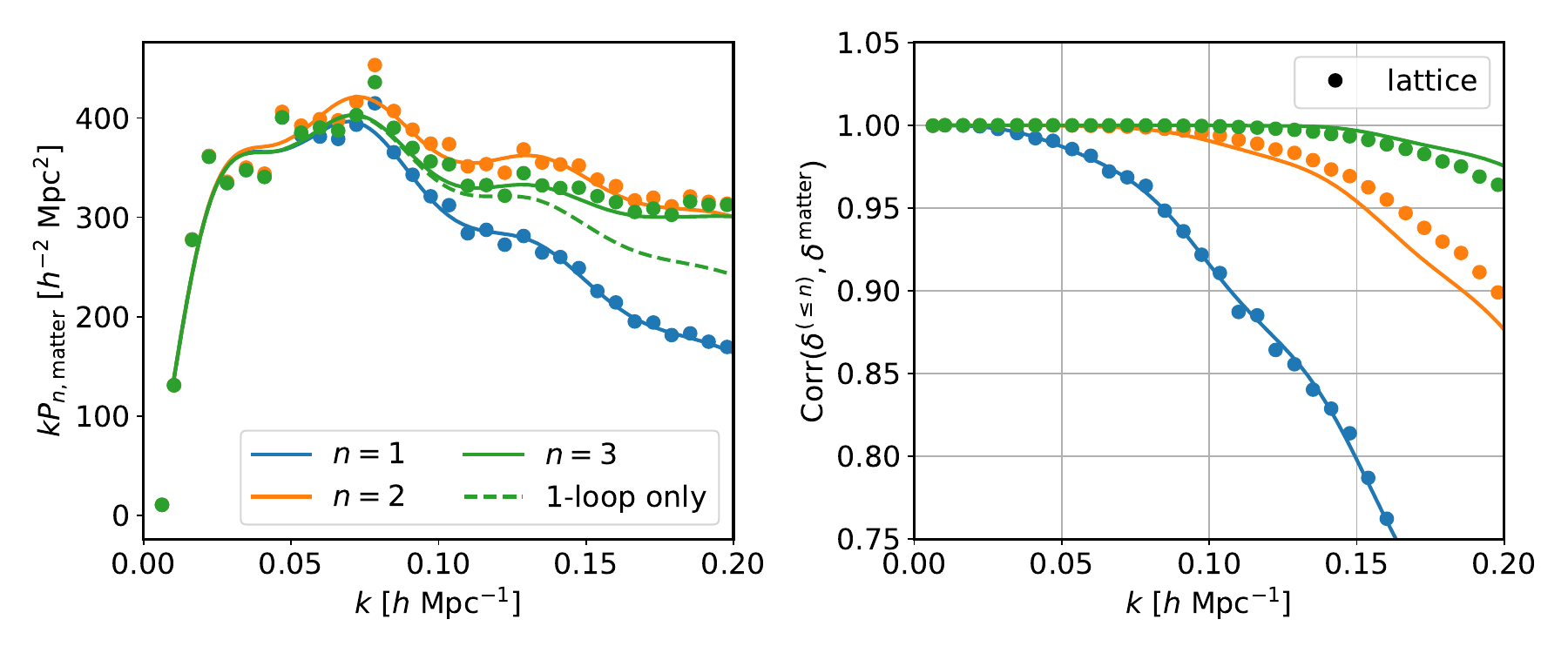}
    \caption{Cross-correlation of the nonlinear and Eulerian matter density fields on the lattice (dots) compared to analytic predictions. (Left) The cross power spectrum between the Eulerian field summed to n$^{\rm th}$ order, computed to 1-loop in perturbation theory, with large-IR contributions at higher loops resummed via an exponential. For the third-order cross correlation there is also an IR-enhanced 2-loop diagram, and we show the prediction without it as a dashed line. (Right) The correlation coefficient between the two fields. The analytic predictions break down around $k \sim 0.15 h$ Mpc$^{-1}$ reflecting the onset of 2-loop mode coupling.}
    \label{fig:matter_corr}
\end{figure}
We are now in a position to better understand the cross correlation of the fully nonlinear and perturbative fields order-by-order. These cross-correlation coefficients are given by
\begin{equation}
    r_{n, {\rm Zel}}(k) = \frac{\langle \delta^{N} | \sum_{i}^n \delta^{(i)} \rangle(k)}{\sqrt{P_{\rm Zel}(k)\sum_{i,j}^{n}\langle \delta^{(i)}| \delta^{(j)}\rangle(k) }}. 
\end{equation}
The right panel of Figure~\ref{fig:zel_corr} shows the measured and predicted correlation coefficients. These are also in excellent agreement until roughly $k \sim 0.15 \ h$ Mpc$^{-1}$, where nonlinear mode coupling departs from the 1-loop prediction. To elucidate these cross-correlation coefficients we can write write $P_{22} = (k\Sigma)^2P_{\rm lin} + P_{\rm 1-loop}$, where $P_{\rm 1-loop}$ is the mode-coupling integral which is left un-cancelled by $2P_{13}$. At one-loop we have $P_{\rm Zel}(k) \approx P_{\rm lin}(k) + P_{\rm 1-loop}(k)$, and we may Taylor expand in the large displacement $k \Sigma$ to have
\begin{align}
    r_{1,\rm Zel}(k) &= \frac{1}{\sqrt{1 + \lambda}}  -  \frac{1}{2\sqrt{1 + \lambda}} k^2 \Sigma^2 + \mathcal{O}(k^4 \Sigma^4), \nonumber \\
    r_{2,\rm Zel}(k) &= 1 - \frac{\lambda}{2(1 + \lambda)} k^2 \Sigma^2 + \mathcal{O}(k^4 \Sigma^4), \nonumber \\
    r_{3, \rm Zel}(k) &= 1 - \mathcal{O}(k^4 \Sigma^4),
\end{align}
where we have defined $\lambda = P_{\rm 1-loop} / P_{\rm lin}$. These expanded correlation coefficients have a few interesting features: First, even though the second order field by itself has an uncancelled IR divergence $k^2 \Sigma^2 P_{\rm lin}$, this divergence is cancelled in the numerator and denominator of the correlation coefficient between the nonlinear and quadratic fields. In the absence of other nonlinearities ($\lambda = 0$), this cancels the leading de-correlation of the two fields, since the IR divergence increases the cross correlation but also the noise in the denominator. Second, differences in the higher-order mode coupling can affect the correlation coefficient as well: in the correlation coefficient between the nonlinear and linear fields, for example, this is almost entirely captured by the factor $1/\sqrt{1+\lambda}$. Since $\lambda < 0$, the lack of mode coupling in the linear field enhances its correlation with the nonlinear field. Similarly, the quadratic field has its leading decorrelation cancelled but retains an order $k^2 \Sigma^2$ decorrelation due to mode coupling, which disappears entirely when cubic operators are added.

The above arguments carry straightforwardly to the cross correlation between the nonlinear matter field and its lattice EPT predictions. Here, Equation~\ref{eqn:zn_recursion}
does not exactly hold, though the infrared contributions derived from it remain the same due to the structure of the EPT mode-coupling kernels \citep{Bernardeau:2001qr}. In this case we can write
\begin{align}
    P_{1,\rm matter}(k) &= \left( (1 + \alpha k^2) P_{\rm lin}(k) + P_{13}(k) + \frac12 k^2 \Sigma^2 P_{\rm lin} \right) e^{-\frac12 k^2 \Sigma^2} \nonumber \\
    P_{2, \rm matter}(k) &= \left( (1 + \alpha k^2) P_{\rm lin}(k) + P_{13}(k) + \frac12 k^2 \Sigma^2 P_{\rm lin} + P_{22}(k) \right)  e^{-\frac12 k^2 \Sigma^2} \nonumber \\
    P_{3, \rm matter}(k) &= \left( (1 + \alpha k^2) P_{\rm lin}(k) + P_{13}(k) + \frac12 k^2 \Sigma^2 P_{\rm lin} + P_{22}(k) \right)  e^{-\frac12 k^2 \Sigma^2}  + P_{3,\rm matter}^{\rm IR-enhanced}
\end{align}
where $\alpha$ is the counterterm fit to the matter autospectrum and $P_{3, \rm matter}^{\rm IR-enhanced}$ is given by swapping the Zeldovich 1-loop terms in Equation~\ref{eqn:p3zel} for the corresponding matter ones. As can be see in the left panel of Figure~\ref{fig:matter_corr}, retaining the IR-enhanced terms beyond 1-loop order is sufficient to obtain very good agreement with the cross spectrum of these two types of fields measured in N-body simulations, with no additional parameters (see \citet{Taruya_2018} for a comparison to the full 2-loop prediction, though without the EFT corrections we have employed here). The right panel of the same figure shows the thus-predicted correlation coefficient, where we have also retained the corresponding 2-loop enhanced IR terms in $P_{33}$, again with very good agreement up to $k \sim 0.15\ h$ Mpc$^{-1}$ where nonlinear mode coupling sets in.

\section{Order-by-order Eulerian bispectra on the lattice}
\label{app:treebk}
In order to measure the tree-level bispectrum from our lattice realizations we create an artificially modulated second-order density field
\begin{equation}
    \delta^{(E)}(\bx; \epsilon) = \delta^{(1)} (\bx) + \epsilon \delta^{(2)}  (\bx) + \epsilon^2 \delta^{(3)}(\bx), 
\end{equation}
where $\epsilon$ is a free parameter and $\delta^{(2)}(\bk)$ is the second-order EPT density field 
\begin{equation}
    \delta^{(2)}(\bx) = \frac{17}{21} \delta^{(1)} (\bx) - \bPsi^{(1)} (\bx) \cdot \nabla \delta^{(1)} (\bx) + \frac{2}{7} s^2 (\bx). 
\end{equation}
The bispectrum of $\delta^{(E)}$ in a given box will be given by the terms
\begin{equation}
    B^{EEE}(k_1, k_2, k_3; \epsilon) = B^{111} + \epsilon (B^{211} + B^{121} + B^{112}) + \epsilon^2 (B^{221} + B^{122} + B^{212} + B^{311} + B^{131} + B^{113}) + \mathcal{O} (\epsilon^3).  
\end{equation}
We know exactly $B^{111}$ for an individual simulation, and thus an estimator for $B^{\rm Tree}$ is given by 
\begin{equation}
    \hat{B}^{\rm Tree} (k_1, k_2, k_3) = \frac{B^{EEE} - B^{111}}{\epsilon}, 
\end{equation}
which is accurate to $\mathcal{O}(\epsilon)$. We use the fiducial value of $\epsilon = 10^{-2}$ in this work, from which we find a tree-level spectrum converged to within $0.1\%$ relative to using $\epsilon = 10^{-3}$. \par 
We can also extract quintic bispectra contributions by considering a pair of quadratic fields $\delta^{(E)}(\bx; \epsilon), \delta^{(E)}(\bx; -\epsilon)$. The combination
\begin{equation*}
 \frac{(B^{EEE}(k_1, k_2, k_3;\epsilon) + B^{EEE}(k_1, k_2, k_3;-\epsilon) ) - 2B^{111}}{2\epsilon^2}
\end{equation*}
estimates the quintic contribution directly. Notice that even though the $\epsilon^2$ terms average to zero, their inclusion as a control variate will contribute to the cross-correlation coefficient. \par 
By modulating higher-order Eulerian contributions with additional powers of $\epsilon$ we maintain strict control over the Eulerian order of the control variate we wish to construct. To extend to $\mathcal{O}(\epsilon^3)$, equivalent to the one-loop matter bispectrum, we would also have to construct the $\delta^{(4)}$ Eulerian field. There is no opposition to doing this in principle -- the recursion relations are known and \cite{Taruya_2018, Taruya:2020qoy} have explored this -- but we elect not to since we do not expect qualitative improvements in Eulerian control variates from extending to one more power in the density field. 
\section{Estimating the Lagrange multiplier $\beta$}
\label{app:betaest}

\begin{figure}
    \centering
    \includegraphics[width=0.7\linewidth]{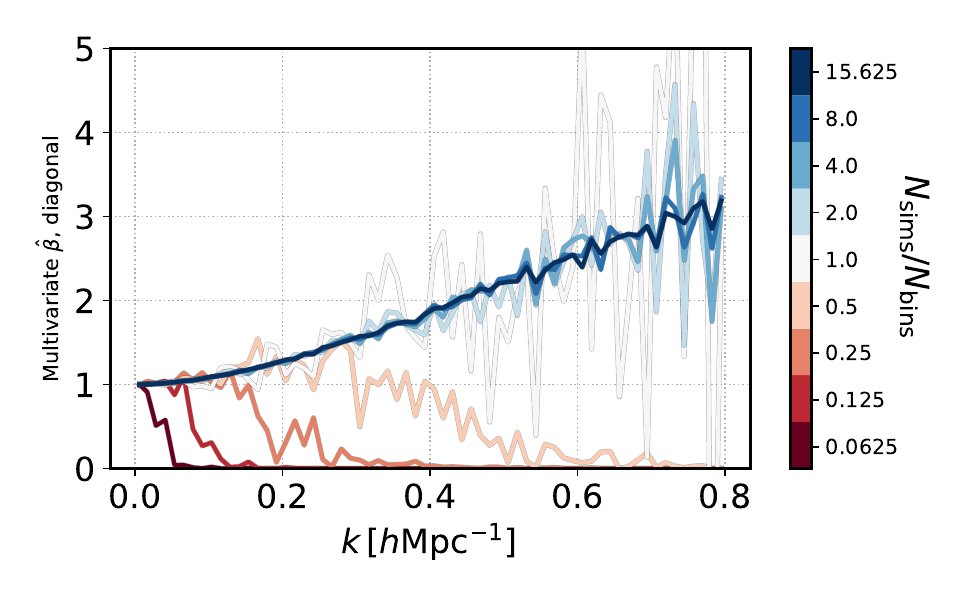}
    \caption{Impact of underdetermination of the control variate covariance matrix on the estimate of $\beta$ due to the use of the Moore-Penrose pseudoinverse. The color indicates how under(over)-determined the covariance matrix estimation is through the parameter $N_{\rm sims} / N_{\rm bins}$. When $N_{\rm sims} / N_{\rm bins} \geq 1$ the control variate covariance is invertible, and we clearly see there is no `damping' of the Lagrange multiplier in question. When the control variate covariance matrix is under-determined and the pseudoinverse is used, a damping is induced which depends on the degree of underdetermination.}
    \label{fig:tanhbust}
\end{figure}

In this appendix we discuss in more detail the suitability of a number of approximations to estimate the Lagrange multiplier, $\beta$, which provides optimal variance reduction. In the univariate problem it is given by (c.f. \cref{eqn:univariatebeta} for the bispectrum-specific case)
\begin{equation}
\label{eqn:unibeta}
    \hat{\beta}_{\rm uni} = \frac{{\rm Cov}(\bx, \boldsymbol{c})}{{\rm diag}({\rm Var} (\boldsymbol{c}))}. 
\end{equation}
In the multi-variate problem (which minimizes the determinant of the covariance matrix of $\boldsymbol{y}$) the equivalent estimate is 
\begin{equation}
\label{eqn:multibeta}
    \hat{\boldsymbol{\beta}}_{\rm multi} = {\rm Cov}(\bx, \boldsymbol{c}) \cdot [{\rm Cov}(\boldsymbol{c})]^{-1}.
\end{equation}
Finally, in \cite{Kokron_2022} we introduced a disconnected approximation to the univeriate Lagrange multiplier, which in the case of the matter power spectrum reduces to 
\begin{equation}
\label{eqn:discbeta}
    \hat{\beta}_{\rm disc} = \left( \frac{P_{N, Z}(k)}{P_N(k)}\right)^2.
\end{equation}
In \cite{Kokron_2022}, the authors checked the validity of \cref{eqn:discbeta} against an estimate of \cref{eqn:multibeta} where the number of simulations $N_{\rm sims}=100$ was smaller than the number of power spectrum bins $N_k = 512$. In order to invert the control variate covariance, the Moore-Penrose pseudoinverse was used to estimate the inverse. A "damping" of the Lagrange multiplier was observed and attributed to being physical in nature, being well fit by a tanh function. \par
This damping is not physical and was spuriously driven by the usage of the pseudoinverse. From the fiducial set of $N=1000$ boxes, we compute the Zeldovich and nonlinear matter $N$-body power spectra at $z=0.5$ in $N_{\rm bins} = 64$, up to $\kmax \sim 0.8\ihmpc$. We then compute the multivariate form of $\beta$ using subsets with $N_{\rm sims} = [4, 8, 16, 32, 64, 128, 256, 512, 1000]$, numerically inverting the Zeldovich power spectrum covariance matrix. In Fig.~\ref{fig:tanhbust} we show the resulting diagonal of the Lagrange multiplier matrix as a function of $N_{\rm sims} / N_{\rm bins}$, the degree of determination of the control variate covariance matrix ${\rm Cov}[\boldsymbol{c}]$. There is clear evidence for the damping being a function of how well-determined this matrix is, and thus a numerical artifact. This damping is a consequence of the construction of the pseudoinverse, where in the singular value decomposition ${\rm Cov}[\boldsymbol{c}] = U\Sigma V$, the diagonal matrix $\Sigma$ containing eigenvalues of ${\rm Cov}[\boldsymbol{c}]$ is inverted. Poorly-determined eigenvalues are set to zero, leading to a matrix with a strictly lower trace than the `true' inverse. The diagonal of the $\beta$ matrix is consequently smaller as a result. The original analysis of \cite{Kokron_2022} was in this underdetermined regime, with $N_{\rm sims} / N_{\rm bins} \sim 0.2$.  \par 
Using this larger sample we now turn to a comparison between the three different estimates \cref{eqn:unibeta,eqn:multibeta,eqn:discbeta} for both the power spectrum and the bispectrum. To compress the dimensionality of the bispectrum analysis we restrict ourselves to equilateral triangles, for which there are $N_{\rm bins} = 23$ in the Zeldovich binning scheme. Additionally, for the bispectrum, the disconnected Gaussian covariance is proportional to $P(k_1)P(k_2)P(k_3)$ and so the appropriate disconnected Lagrange multiplier is 
 \begin{equation}
\label{eqn:discbetabk}
    \hat{\beta}_{\rm disc} (k_1, k_2, k_3) = \frac{P_{N, Z}(k_1)P_{N, Z}(k_2)P_{N, Z}(k_3)}{P_N(k_1)P_N(k_2)P_N(k_3)}. 
\end{equation}
\begin{figure*}
    \centering
    \includegraphics[width=\linewidth]{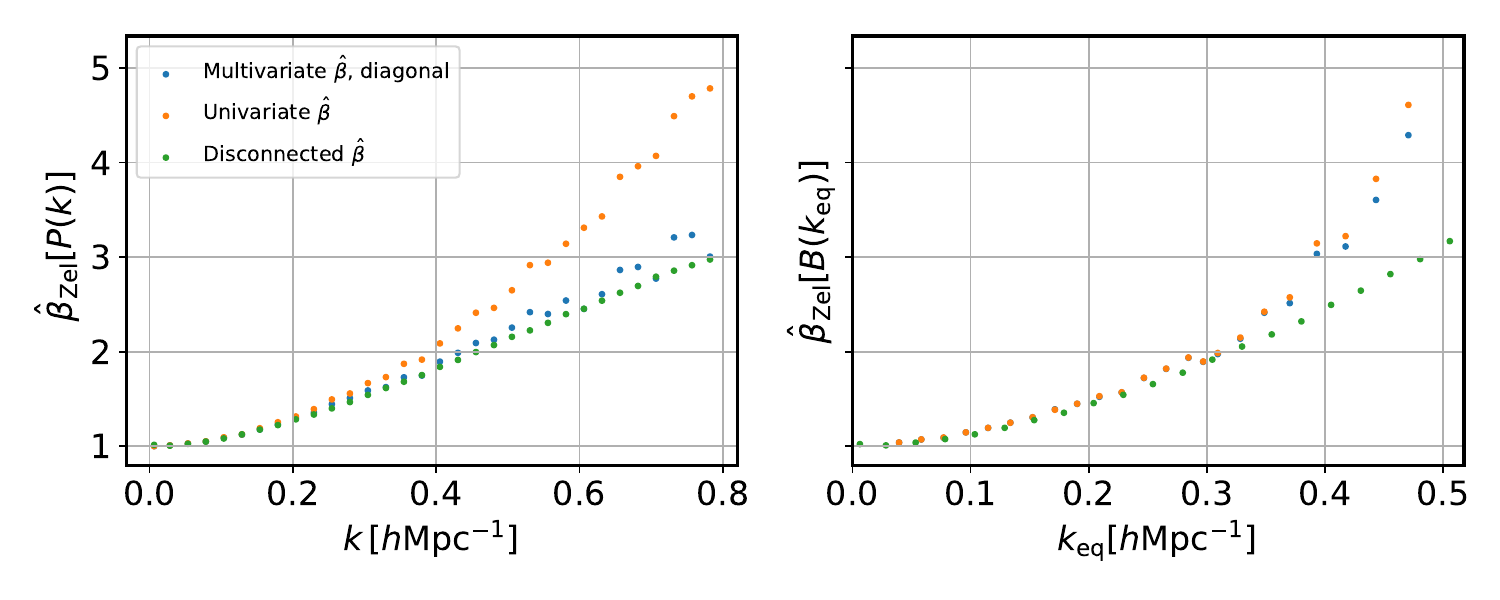}
    \caption{Estimate of the Lagrange multiplier, $\hat{\beta}$ for the Zeldovich control variate in the case of the power spectrum (left) and bispectrum (right), for equilateral triangle configurations. The blue points show the diagonal contribution from the full multivariate estimate of $\beta$, the orange points show the univariate estimate from the bin-by-bin problem, and the green points show the disconnected approximation to the Lagrange multiplier.}
    \label{fig:betazeldovich}
\end{figure*}
Fig.~\ref{fig:betazeldovich} shows the three different estimates for $\beta$. In the left panel, which shows estimates for the power spectrum, we can see that all three estimates agree well until $k\sim 0.2 \ihmpc$ at which point they begin to diverge. The disconnected approximation closely tracks the diagonal of the multivariate term through all scales considered, while the univariate $\hat{\beta}$ grows at a faster rate. We stress there is no reason a priori that these different estimates of $\beta$ should closely agree across all scales -- the multivariate control variate problem optimizes over a fundamentally different objective function than the univariate bin-by-bin problem. In the case of the power spectrum we see the disconnected approximation is a close match to the multivariate estimator. In the case of the equilateral bispectrum the picture is somewhat different -- the univariate and multivariate estimates of $\beta$ agree closely for all triangles considered, while the disconnected approximation underestimates the covariance-based values of $\beta$ starting at $k_{\rm equi}=0.3 \ihmpc$. This can be understood by studying the behavior of the bispectrum precision, shown in Fig.~\ref{fig:bsnr}. $k\approx 0.3 \ihmpc$ is precisely the triangle scale at which the SNR estimated from using the disconnected nonlinear covariance begins to disagree from the full empirical estimate -- a proxy for the importance of connected terms in the covariance. \par 
How much variance reduction is lost from adopting a slightly sub-optimal value of $\beta$? Suppose one uses $\beta = \beta^* (1 + \epsilon)$ where $\beta^*$ is the optimal multiplier for the univariate control variate problem. In this case, a straightforward calculation shows the variance reduction goes to 
\begin{equation}
    \frac{\sigma_y^2}{\sigma_x^2} = (1 - \rho^2) + \epsilon^2 \rho^2.
\end{equation}
For the disconnected approximations shown in Fig.~\ref{fig:betazeldovich}, we find that the largest amount observed is $\epsilon^2 \approx 0.15$ at the smallest scales considered, where the cross-correlation coefficient $\rho$ is already suppressed.

\end{appendix}
\bibliographystyle{apsrev4-1}

\bibliography{main}

\end{document}